\documentclass[12pt]{article}

\ifx\pdfoutput\undefined
\usepackage[dvips,bookmarks]{hyperref}
\else
\usepackage{hyperref}
\fi
\hypersetup{colorlinks=false,bookmarksopen,bookmarksnumbered,citecolor=blue,
   pdfstartview=FitH}

\usepackage[pdftex]{graphicx}	
\usepackage{latexsym}

\usepackage{amssymb,amsfonts,amsmath}
\usepackage{graphicx} 
\usepackage{indentfirst}

 \usepackage{bbm}

\parskip=\medskipamount

\newcommand {\cD}{{\cal D}}
\newcommand {\cE}{{\cal E}}

\newcommand {\cJ}{{\cal J}}
\newcommand {\cK}{{\cal K}}
\newcommand {\cL}{{\cal L}}
\newcommand {\cM}{{\cal M}}
\newcommand {\cN}{{\cal N}}
\newcommand {\cO}{{\cal O}}

\newcommand {\cS}{{\cal S}}

\newcommand {\cV}{{\cal V}}

\newcommand {\cX}{{\cal X}}

\newcommand {\cZ}{{\cal Z}}

\newcommand{\bL}{{\bf L}}

\newcommand{\bR}{{\bf R}}

\def\a{\alpha}

\def\b{\beta}

\def\d{\delta}
\def\e{\epsilon}

\def\g{\gamma}

\def\j{\psi}

\def\l{\lambda}

\def\o{\omega}
\def\p{\pi}
\def\q{\theta}
\def\r{\rho}
\def\s{\sigma}
\def\t{\tau}

\def\x{\xi}
\def\z{\zeta}
\def\D{\Delta}
\def\F{\Phi}
\def\Fb{{\bar{\Phi}}}

\def\L{\Lambda}
\def\O{\Omega}

\def\U{\Upsilon}

\def\rd{{\rm d}}
\def\ri{{\rm i}}
\def\re{{\rm e}}

\newcommand{\ve}{\varepsilon}                            
\newcommand{\cDB}{{\bar\cD}}                            

\newcommand{\pa}{\partial}                           
\newcommand{\hf}{\frac12}

\newcommand{\be}{\begin{equation}}
\newcommand{\ee}{\end{equation}}
\newcommand{\bea}{\begin{eqnarray}}
\newcommand{\eea}{\end{eqnarray}}
\newcommand{\non}{\nonumber}
\newcommand{\1}{{\underline{1}}}
\newcommand{\2}{{\underline{2}}}

\newcommand{\bm}[1]{\mbox{\boldmath$#1$}}

\def\double #1{#1{\hbox{\kern-2pt $#1$}}}

\newcommand{\qb}{{\bar{\theta}}}

\newif\ifdtup

\newcommand{\bsubeq}{\begin{subequations}}
\newcommand{\esubeq}{\end{subequations}}

\newcommand{\bai}{{\bar i}}
\newcommand{\baj}{{\bar j}}
\newcommand{\bak}{{\bar k}}
\newcommand{\bal}{{\bar l}}

\newcommand{\bau}{{\bar 1}}
\newcommand{\bad}{{\bar 2}}

\newcommand{\rL}{{\rm L}}
\newcommand{\rR}{{\rm R}}

\newcommand{\veps}{\varepsilon}

\numberwithin{equation}{section}

\begin{document}
\begin{titlepage}
\begin{flushright}
February, 2014\\
Revised version: March, 2014\\
\end{flushright}
\vspace{5mm}

\begin{center}
{\Large \bf 
$\bm{\cN=4}$ supersymmetric Yang-Mills theories in AdS$_3$}\\ 
\end{center}

\begin{center}

{\bf
Sergei M. Kuzenko
and
Gabriele Tartaglino-Mazzucchelli
} \\
\vspace{5mm}

\footnotesize{
{\it School of Physics M013, The University of Western Australia\\
35 Stirling Highway, Crawley W.A. 6009, Australia}}  
~\\
\texttt{gabriele.tartaglino-mazzucchelli@uwa.edu.au}
\vspace{2mm}

\end{center}

\begin{abstract}
\baselineskip=14pt
For all types of $\cN=4$ anti-de Sitter (AdS) supersymmetry in three dimensions, 
we construct manifestly supersymmetric actions for Abelian vector multiplets
and explain how to extend the construction to the non-Abelian case.
Manifestly $\cN=4$ supersymmetric Yang-Mills (SYM) actions 
are explicitly given in the cases of (2,2) and critical (4,0) AdS supersymmetries.
The $\cN=4$ vector multiplets and the corresponding actions are then reduced 
to (2,0) AdS superspace, in which only $\cN=2$ supersymmetry is manifest. 
Using the off-shell structure of the $\cN=4$ vector multiplets, 
we provide complete $\cN=4$ SYM actions in (2,0) AdS superspace
for all types of $\cN=4$ AdS supersymmetry. 
In the case of (4,0) AdS supersymmetry, which admits a Euclidean counterpart, 
the resulting $\cN=2$ action 
contains a Chern-Simons term proportional to $\bm q /r$, 
where $r$ is the radius of AdS$_3$
and $\bm q$ is the $R$-charge of a chiral scalar superfield.  
The $R$-charge is a linear inhomogeneous function of $X$, 
an expectation value of the $\cN=4$ Cotton superfield. 
Thus our results explain 
the mysterious structure of $\cN=4$ supersymmetric Yang-Mills theories on $S^3$ 
discovered in  arXiv:1401.7952. 
In the case of (3,1) AdS supersymmetry, 
which has no Euclidean counterpart, the SYM action contains both 
a Chern-Simons term 
and a chiral mass-like term.
In the case of (2,2) AdS supersymmetry, which admits a Euclidean counterpart,
the SYM action has no Chern-Simons and chiral mass-like terms. 
\end{abstract}

\vfill

\vfill
\end{titlepage}

\newpage
\renewcommand{\thefootnote}{\arabic{footnote}}
\setcounter{footnote}{0}

\tableofcontents{}
\vspace{1cm}
\bigskip\hrule


\section{Introduction}
\setcounter{equation}{0}

Recently, Samsonov and Sorokin \cite{SS} have constructed $\cN=4$ supersymmetric 
Yang-Mills (SYM) theories on $S^3$, both in terms of $\cN=2$ superfields and component fields. 
In the $\cN=2$ superspace setting, such a theory describes coupling of the vector multiplet 
to a chiral scalar multiplet of $R$-charge $\bm q$, with  $\bm q$ arbitrary.\footnote{As argued in \cite{SS},
a natural bound on the values of the $R$-charge emerges, $0\leq \bm q \leq 2$, 
if the spectrum of the theory is required to be free of negative energy states.} 
The case $\bm q=1$ was actually considered
at the  component level three years earlier by Hama, Hosomichi and Lee \cite{HHL}, 
although the structure of extended supersymmetry transformations was not clarified
by these authors. The remarkable feature of the $\cN=4$ SYM theories on $S^3$ 
constructed in \cite{SS,HHL} is the fact that the $\cN=4$ supersymmetry 
requires the action to include a Chern-Simons term
proportional to $\bm q /r$, where $r$ denotes the radius of $S^3$. 
The presence of such a Chern-Simons term calls for an explanation within 
a manifestly $\cN=4$ supersymmetric formulation of the theory. 

Supersymmetric theories on $S^3$ may naturally be obtained from those defined on 
three-dimensional (3D) anti-de Sitter space, AdS$_3$, 
by Wick rotation.\footnote{Since $S^3$ and AdS$_3$ have different topologies, 
Wick rotation is a rather formal procedure, and some additional care is required in order 
to make it well defined.} 
There are three types of 3D $\cN=4$ AdS superspaces 
\cite{KLT-M12}, in accordance with the existence of several versions of 
$\cN$-extended AdS  supergravity in three dimensions, 
the $(p,q)$  AdS supergravity theories \cite{AT}, where $p+q = \cN$ and $p \geq q$.
These are the (4,0), (3,1) and (2,2) AdS superspaces. 
Furthermore, 
there exist three inequivalent versions 
of (4,0) AdS superspace \cite{KLT-M12}:
\begin{subequations}
\bea 
X &=& 0 ~; \label{1.1a}  \\
X & \neq & 0~, \qquad |X|\neq 2S ~;   \label{1.1b}  \\
|X|&=&  2S ~.
\label{1.1c} 
\eea
\end{subequations}
Here the paramaters $X$ and $S$ 
are constant AdS values of the $\cN=4$ Cotton superfield and  
one of the superspace torsion components respectively 
\cite{HIPT,KLT-M11}.\footnote{It is natural 
to think of $S$ as a superspace analogue of the square root of the scalar curvature.  However, there is no obvious 
spacetime interpretation for $X$.}
The Cotton superfield automatically vanishes, 
$X=0$, for the  (3,1) and (2,2) AdS superspaces \cite{KLT-M12}. 
The $\cN=4$ AdS superspaces are conformally flat if and only if $X=0$ \cite{KLT-M12}. 
The (3,1) AdS superspace has no Euclidean analogue. 

Building on the off-shell formulation for general 3D $\cN=4$ supergravity-matter systems 
given in \cite{KLT-M11}, Ref.~\cite{BKT-M} provided a powerful formalism\footnote{This
formalism is inspired by  the projective-superspace approach to 4D 
$\cN=2$ supersymmetric theories in Minkowski space \cite{KLR,LR1,LR2}.}  
(off-shell supermultiplets, manifestly supersymmetric action principles etc.)
to construct off-shell supersymmetric theories in all the $\cN=4$ AdS superspaces, 
as well as to reduce these theories to $\cN=2$ AdS superspaces \cite{KT-M11}.  
However, the analysis in \cite{BKT-M}  was restricted to the case of the most general $\cN=4$ 
supersymmetric nonlinear sigma models, due to their remarkably rich 
geometric structure
and diverse physical properties associated with the different types of $\cN=4$ AdS
supersymmetry.  In the present note we apply the formalism of \cite{BKT-M} to construct 
 SYM theories in the $\cN=4 $ AdS superspaces. 

This paper is organized as follows. In section 2 we review the geometry of 
the various 3D $\cN=4$ AdS superspaces. In section 3 we review 
the main results concerning the $\cN=4$ vector multiplets in conformal supergravity 
and construct a new family of composite linear multiplets in $\cN=4$ AdS superspaces.   
The latter result is used in section 4 to construct the $\cN=4$ vector multiplet actions in 
all the $\cN=4$ AdS superspaces. Sections 5 and 6 are devoted to the reduction 
of the results obtained to (2,0) AdS superspace. Concluding comments are given 
in section 7. The paper contains five technical appendices. In appendix A we discuss
the projective-superspace formulation for a 3D $\cN=4$ Yang-Mills multiplet 
in conformal supergravity. The isometries of $\cN=4$ AdS superspaces are reviewed 
in appendix B. The fundamentals of (2,0) AdS superspace are collected in appendix C.
 In  appendix D  we present complete $\cN=4$ SYM actions 
in (2,0) AdS superspace for all types of $\cN=4$ AdS supersymmetry. 
Finally, in appendix E we relate the $\cN=4$ and $\cN=2$ superspace formulations 
for $\cN=4$ SYM theories in AdS$_3$. 


\section{$\cN=4$ AdS superspaces}
\setcounter{equation}{0}
\label{N4AdSsuperspaces}

In this section we review the salient points of the geometry of the various $\cN=4$ AdS  superspaces constructed 
in \cite{KLT-M12}. 

According to the on-shell supergravity analysis of \cite{AT}, 
there are three types of $\cN=4$ AdS supersymmetry in three dimensions.
This implies the existence of  
three inequivalent maximally symmetric and conformally flat $(p,q)$ AdS superspaces
\bea
{\rm AdS}_{(3|p,q)} = \frac{ {\rm OSp} (p|2; {\mathbb R} ) \times  {\rm OSp} (q|2; {\mathbb R} ) } 
{ {\rm SL}( 2, {\mathbb R}) \times {\rm SO}(p) \times {\rm SO}(q)}~, \qquad
p+q =4~, \qquad p\geq q~.
\eea
In accordance with  the more recent analysis of \cite{KLT-M12}, 
which was  based on the use  of the off-shell formulation for 3D $\cN=4$ 
conformal supergravity \cite{HIPT,KLT-M11}, 
there exist two more inequivalent versions of (4,0) AdS superspace. These superspaces 
are not conformally flat and correspond to the choices \eqref{1.1b} and \eqref{1.1c}.
Their existence is due to the fact that for $\cN \geq 4$ there exist  more general 
AdS supergroups in the case $p -\cN = q=0$,  
than those considered by Ach\'ucarro and Townsend \cite{AT}.

All the $\cN=4$ AdS superspace geometries may be described using  covariant derivatives
of the general form: 
\bea
\cD_A=(\cD_a,\cD_\a^{i\bai})=E_A{}^M\pa_M+\hf\O_A{}^{cd}\cM_{cd}
+\F_A{}^{kl}\bL_{kl}
+\F_A{}^{\bak\bal}\bR_{\bak\bal}
~.
\label{N4-dev}
\eea
Here the operators $\bL_{kl}$ and $\bR_{\bak\bal}$ generate the $R$-symmetry group
SU(2)$_\rL\times$SU(2)$_\rR$ and act on the covariant derivatives as
\bea 
[\bL^{kl},\cD_\a^{i\bai}]=\ve^{i(k}\cD_\a^{l)\bai}
~,~~~~~~
[\bR^{\bak\bal},\cD_\a^{i\bai}]=\ve^{\bai(\bak}\cD_\a^{i\bal)}
~.
\eea
For each of the $\cN=4$ AdS superspaces, the covariant derivatives obey (anti-)commutation 
relations of the form \cite{KLT-M12}:
\bsubeq \label{N=4alg}
\bea
\{\cD_\a^{i\bai},\cD_\b^{j\baj }\}&=&\phantom{+}
2\ri\,\ve^{ij}\ve^{\bai \baj }\cD_{\a\b}
-\,4\ri(\cS^{ij}{}^{\bai \baj }+\ve^{ij}\ve^{\bai \baj }\cS)\cM_{\a\b}
\non\\
&&
+\,{2\ri}\ve_{\a\b}\ve^{\bai \baj }(2\cS+X)\bL^{ij}
-\,2\ri\ve_{\a\b}\ve^{ij}\cS^{kl}{}^{\bai \baj }\bL_{kl}
\non\\
&&
+\,2\ri\ve_{\a\b}\ve^{ij}(2\cS-X)\bR^{\bai\baj}
-\,2\ri\ve_{\a\b}\ve^{\bai \baj }\cS^{ij}{}^{\bak\bal}\bR_{\bak\bal}
~,
\label{N=4alg-1}
\\
{[}\cD_{\a\b},\cD_\g^{k\bak}{]}&=&
-\,2\Big(\d^k_l\d^\bak_\bal\cS+\cS^k{}_l{}^\bak{}_\bal\Big)\ve_{\g(\a}\cD_{\b)}^{l\bal}
~,
\label{N=4alg-3/2}
\\
{[}\cD_a,\cD_b{]}&=&-\,4\,S^2\cM_{ab}~,
\label{N=4alg-2}
\eea
\esubeq
where the real tensor  $\cS^{ij\bai\baj} = \cS^{(ij)(\bai\baj)}$ is covariantly constant, 
and the real scalars $\cS$, $X$ and $S$ are constant.
The parameter $S$ determines the curvature of AdS$_3$.
Depending on the  superspace type, the 
parameters  $\cS$, $\cS^{ij\bai\baj}$ and $X$ have the  following explicit form \cite{KLT-M12}:
\bsubeq \label{SSX--}
\bea
&\text{(4,0) AdS}:&~~~\cS=S~,~~~\cS^{ij\bai\baj}=0~,~~~~~~
X{~\rm arbitrary}~;
\label{SSX40}
\\
&\text{(3,1) AdS}:&~~~\cS=\hf S~,~~
\cS^{ij\bai\baj}=
\hf 
\Big(\ve^{ij}\ve^{\bai\baj}-2w^{i\bai}w^{j\baj}\Big)S
~,~~~
X=0~;
\label{SSX31}
\\
&\text{(2,2) AdS}:&~~~\cS=0~,~~~
\cS^{ij\bai\baj}=l^{ij}r^{\bai\baj}\,S
~,~~~X=0~.
\label{SSX22}
\eea
\esubeq
In the (3,1) case, the covariantly constant tensor $w^{i\bai}$ is real, 
$\overline{w^{i\bai}} = w_{i\bai}= \ve_{ij}\ve_{\bai \baj} w^{j\baj}$, and normalized as 
\bea
w^{i\bak}w_{i\bak}=\d^i{}_j~, \qquad w^{k\bai}w_{k\baj}= \d^{\bai}{}_{\baj}~.
\eea
In the (2,2) case, the real iso-triplets $l^{ij} = l^{ji}$ and $r^{\bai\baj}=r^{\baj\bai}$ 
are covariantly constant and normalized as
\bea
l^{ik}l_{kj} = \d^i{}_j~, \qquad r^{\bai\bak}r_{\bak\baj}=\d^{\bai}{}_{\baj}~.
\eea

We emphasize that $X$ 
can appear in the algebra only in the (4,0) case. 
For general values of $X$, the tangent space group of the (4,0) AdS supergeometry is the full 
$R$-symmetry group SU(2)$_\rL\times$SU(2)$_\rR$.
For the two critical values, $X=2S$ and $X=-2S$,
the SU(2)$_\rR$ or SU(2)$_\rL$ group, respectively, can be gauged away.

In the non-critical case, $|X|\neq 2S$, the isometry group of (4,0) AdS 
superspace is isomorphic to\footnote{We are grateful to Igor Samsonov 
and Dima Sorokin for this observation.} 
\bea
{\rm D} (2, 1;\a) \times {\rm SL}(2,{\mathbb R})~,\qquad \a \neq -1, 0~,
\eea
for some $\a \in \mathbb R$.  
Here ${\rm D} (2, 1;\a)$ is one of the exceptional simple supergroups, 
see, e.g., \cite{DeWitt} for a review.\footnote{Various supercosets based on the 
exceptional supergroup   ${\rm D} (2, 1;\a)$ were considered in \cite{BILS}.}
As is known, not all values of the real parameter $\a$ 
lead to distinct supergroups. The point is that  
there is a finite group $G$ 
(of order 6) of fractional linear transformations 
of ${\mathbb R}P^1 ={\mathbb R} \cup \{ \infty \}$, 
the compactified real line, with the property 
that any transformation $g \in G$ maps $\a \to \a '  = g(\a)$ such that 
${\rm D} (2, 1;\a)$ and ${\rm D} (2, 1;\a ')$ are isomorphic \cite{DeWitt}.
The subset $\{ -1, 0 , \infty \}$ proves to be  fixed under the action of $G$. 
Up to an isomorphism,  it suffices to restrict $\a$  to the range $0<\a \leq 1$.
The case $\a =1$  corresponds to the conformally flat (4,0) 
AdS superspace, for which $X=0$. The isometry group of this superspace is 
${\rm OSp}(4|2) \times {\rm SL}(2,{\mathbb R})$.
In general, there is a correspondence between $\a$ and the (4,0)
AdS parameter ${\bm q} = 1 +\frac{X}{2S}$, which will  play an important role in 
this paper.
These parameters may be identified in the domain $0<\a \leq 1$.
 The choice $\a =0$  corresponds to ${\bm q} =0$,
which is  one of the two critical  (4,0) AdS cases.\footnote{The two critical
(4,0) AdS superspaces,  which are characterized by the choices $\bm q =0$ and 
$\bm q =2$, have isomorphic isometry groups.}
The isometry group of this (4,0) AdS superspace degenerates to 
${\rm SU} (1, 1|2) \times {\rm SL}(2,{\mathbb R})$, see also the discussion in \cite{BILS}.

For the (3,1) and (2,2) AdS geometries, the $R$-symmetry sector of
the  superspace holonomy group is a subgroup of
SU(2)$_\rL\times$SU(2)$_\rR$ 
\cite{KLT-M12}.
For the (3,1) supergeometry, the relevant subgroup is SU(2)$_\cJ$ generated by
\bea
\cJ_{kl}=\bL_{kl}+w_k{}^\bak w_l{}^\bal\bR_{\bak\bal}
~,~~~{\rm or}~~~
\cJ_{\bak\bal}=w^k{}_\bak w^l{}_\bal\bL_{kl}+\bR_{\bak\bal}=w^k{}_\bak w^l{}_\bal\cJ_{kl}
~.
\eea
The generators $\cJ_{kl}$ and $\cJ_{\bak\bal}$ leave $w^{i\bai}$ invariant, 
 $\cJ_{kl}w^{i\bai}= \cJ_{\bak\bal} w^{i\bai}= 0$.
Since the $R$-symmetry curvature is spanned by the generators of  SU(2)$_\cJ$,  
it is possible to choose a gauge in which 
the $R$ symmetry connection takes its values in the Lie algebra of  SU(2)$_\cJ$;
in this gauge, the parameter $w^{i\bai}$ is constant.

In the (2,2) case,
 the $R$-symmetry sector of
the  superspace holonomy group is  
the Abelian subgroup $\rm U(1)_\rL \times U(1)_\rR $ of $\rm SU(2)_\rL \times SU(2)_\rR$    
generated by 
\bea
\bL:=l^{kl}\bL_{kl}
~,~~~~~~
\bR:=r^{\bak\bal}\bR_{\bak\bal}~.
\eea
This subgroup leaves invariant  the covariantly constant parameters $l^{kl}$ and  $r^{\bak\bal}$.
In the remainder of the paper, we choose a gauge in which only 
this subgroup 
appears in the (2,2) covariant derivatives. In this gauge the parameters
$l^{kl}$ and $r^{\bak\bal}$ are constant.

\section{Vector multiplets in $\cN=4$ AdS superspaces}
\setcounter{equation}{0}

There are two inequivalent  $\cN=4$ vector multiplets in three dimensions, 
left and right ones.\footnote{The existence of two inequivalent 3D $\cN=4$ vector multiplets
was first discussed by Brooks and Gates \cite{BrooksG}. 
The modern off-shell formulation for these multiplets was given by Zupnik \cite{Zupnik99}
in the rigid supersymmetric case. In the locally supersymmetric case, these multiplets were 
described in \cite{KLT-M11}.} 
 In a curved  $\cN=4$ superspace \cite{KLT-M11}, they may be  described
 in terms of  gauge-invariant field strengths, 
 $W^{ij}=W^{ji} = \overline{W_{ij}} $ and $W^{\bai\baj}=W^{\baj\bai} = \overline{W_{\bai\baj}}$, 
 which transform under the left and right subgroups of  
 the supergravity $R$-symmetry group $\rm SU(2)_\rL \times SU(2)_\rR$, respectively, 
 and obey  the {\it inequivalent } analyticity constraints\footnote{Here we focus our attention on the Abelian vector 
 multiplets. In appendix \ref{AppendixA} we elaborate on the non-Abelian case.}
\bsubeq
\bea
\cD_\a^{(i\bai}W^{kl)}&=&0
~,
\label{a0-L}
\\
\cD_\a^{i(\bai}W^{\bak\bal)}&=&0
~.
\label{a0-R}
\eea
\esubeq
A real symmetric isospinor $W^{ij}$ under the constraint (\ref{a0-L}) is called 
a left linear multiplet.
Similarly, eq. (\ref{a0-R}) defines a right linear multiplet.

The field strengths introduced may be interpreted as special examples 
of the covariant projective $\cN=4$ supermultiplets studied in \cite{KLT-M11}. 
Let us introduce left and right isospinor variables, 
$v_\rL:=v^i \in {\mathbb C}^2 \setminus  \{0\}$ and 
$v_\rR:=v^\bai \in {\mathbb C}^2 \setminus  \{0\}$, 
and use them to define two different subsets,  $\cD_\a^{(1)\bai}$ and $\cD_\a^{(\bau)i}$,
in the set of spinor covariant  derivatives $\cD^{i\bar i}_\a$,  
\bea
\cD_\a^{(1)\bai}:=v_i\cD_\a^{i\bai}~,~~~~~~
\cD_\a^{(\bau)i}:=v_\bai\cD_\a^{i\bai}~,
\label{3.2}
\eea
as well as the index-free superfields
\bea
W_\rL^{(2)}:=v_iv_jW^{ij}\equiv W^{(2)}~,~~~~~~
W_\rR^{(2)}:=v_\bai v_\baj W^{\bai\baj}\equiv W^{(\bad)}
\eea
associated with the left and the right linear multiplets, respectively.
Now, the constraints (\ref{a0-L}) and (\ref{a0-R}) turn into the 
generalized chirality conditions 
\begin{subequations}
\bea
\cD_\a^{(1)\bai}W_\rL^{(2)}&=&0~, 
\label{6.24a} \\
\cD_\a^{(\bau)i}W_\rR^{(2)}&=&0~.
\label{6.24b} 
\eea
\end{subequations}
The superfield $W_\rL^{(2)} (v_\rL)$ is called a left $\cO(2)$ multiplet. 
Similarly, $W_\rR^{(2)} (v_\rR)$ is called a right $\cO(2)$ multiplet. 

All results concerning left vector multiplets may be related to the right ones
by applying the so-called mirror map \cite{Zupnik99,KLT-M11}. Therefore we restrict 
our analysis to the case of left vector multiplets.

As shown in \cite{KLT-M11}, 
the constraint (\ref{a0-L}) may be solved in terms of an unconstrained gauge prepotential
that is  a right weight-zero  tropical multiplet $V_\rR(v_\rR)$.
The most general solution to the analyticity constraint (\ref{a0-L})  is
\bea
W^{ij}
&=&
\frac{\ri}{4}\big(
\cD^{ij \bai  \baj}
-4\ri\cS^{ij\bai\baj}\big)
\oint_\g \frac{(v_\rR, \rd v_\rR)}{2\pi}
\frac{u_\bai u_\baj}{(v_\rR ,u_\rR)^2}
V_\rR(v_\rR) ~, 
\qquad (v_\rR,  u_\rR) := v{}^{\bar i}u_{\bar i}
~,~~~
\label{3.5}
\eea
where we have defined 
\bea
&&\cD^{ij \bai  \baj}:=\cD^{\a (i ( \bai}\cD_\a^{j)\baj)}
~.~~~~~~
\eea
The right-hand side of (\ref{3.5}) involves a constant isospinor $u_\rR = u^{\bar i}$ 
constrained only by the  condition $(v_\rR,  u_\rR) \neq 0$ 
which must hold along the closed integration contour $\g$. 
It can be shown that (\ref{3.5}) is invariant under an arbitrary infinitesimal 
variation of $u_\rR$, which may be represented as  
 $\d u_\rR = \a u_\rR + \b v_\rR $, with $\a, \b \in {\mathbb C}$.
Thus $W^{ij}$ is independent of $u_\rR$.
The right-hand side of \eqref{3.5} is invariant under gauge transformations 
\bea
\d V_\rR = \ri ( \breve{\l}_\rR -\l_\rR) ~, 
\eea
where the gauge parameter $\l_\rR (v_\rR) $ is a right arctic weight-zero multiplet, 
see \cite{KLT-M11} for more details. 

It is important to point out that the field strength of the left vector multiplet, 
$W_\rL^{(2)} (v_\rL)$, is a left projective multiplet. 
However, its gauge prepotential,  $V_\rR(v_\rR)$, is a right projective multiplet.

All previous results in this section hold for the general curved $\cN=4$ superspace
as defined in \cite{KLT-M11}.
The specific feature of  the AdS geometries is that 
 a composite right $\cO(2)$ multiplet may be constructed 
 starting from  the left vector multiplet.
Consider the tensor superfield
\bea
{\bm W}^{\bai\baj}
:=
-\frac{\ri}{12}\cD^{ij\bai\baj}W_{ij}~, 
\label{3.8}
\eea
which can equivalently be realized as 
the right isotwistor superfield 
\bea
{\bm W}^{({2})}_\rR (v_\rR):=v_\bai v_\baj {\bm W}^{\bai\baj} 
\equiv {\bm W}^{({\bar 2})}~.
\label{3.8b}
\eea
Making use of the algebra
\bsubeq
\bea
\{\cD_\a^{(\bar 1)i},\cD_\b^{(\bar 1)j}\}&=&
-\,4\ri\cS^{(\bar{2})}{}^{ij}\cM_{\a\b}
-\,2\ri\ve_{\a\b}\ve^{ij}\cS^{(\bar{2})}{}^{kl}\bL_{kl}
+\,2\ri\ve_{\a\b}\ve^{ij}(2\cS-X)\bR^{(\bar{2})}
~, ~~~~~\label{3.9a}
\\
&&\cS^{(\bar{2})}{}^{ij}:=v_\bai v_\baj \cS^{ij}{}^{\bai\baj}
~,~~~~~~
{\bR}^{(\bar{2})}:=v_\bai v_\baj {\bR}^{\bai\baj}
\eea
\esubeq
in conjunction with the equations
\bea
&&
{[}\bR^{(\bar{2})},\cD_\a^{(\bar{1})i}{]}=0
~,~~~~~~
\bR^{(\bar{2})}W^{ij}=0
~,
~~~~~~
\cD_\a^{i\bai}W^{jk}
=
\frac{2}{3}\ve^{i(j}\cD_{\a l}^{\bai}W^{k)l}
~,
\eea
it is a short calculation to prove that 
\bea
\cD_\a^{(\bar 1)i}{\bm W}^{({2})}_\rR=0
~~~~~~\Longleftrightarrow
~~~~~~
\cD_\a^{i(\bai}{\bm W}^{\baj\bak)}
=0~.
\eea
Therefore, ${\bm W}^{\bai\baj}$ is a right linear multiplet.\footnote{Starting from
${\bm W}^{\bai\baj}$, we can construct a left linear multiplet, and so on and so forth. 
As a result, we have a procedure to generate higher-derivative left and right
linear multiplets.} This superfield and its mirror image 
will be our crucial building blocks to construct $\cN=4$ SYM actions in AdS$_3$.

It is possible to express
${\bm W}^{({2})}_\rR$ in terms of the gauge prepotential $V_\rR$.
The result is
\bea
{\bm W}^{({2})}_\rR(v_\rR)
&=&
\D^{(4)}_\rR
\oint \frac{(\hat{v}_\rR, \rd \hat{v}_\rR)}{2\p(v_\rR,\hat{v}_\rR)^2}
V_\rR(\hat{v}_\rR)
~.~~~~~~~~~
\label{WD4V}
\eea
Here we have introduced the right analyticity projection operator
\bea
\D^{(4)}_\rR&=&
\frac{1}{48}
\cD^{(\bad)kl}\big(
\cD^{(\bad)}_{kl}
-4\ri \cS^{(\bad)}_{kl}
\big)
=
\frac{1}{48}
\big(
\cD^{(\bad)kl}
-4\ri \cS^{(\bad)kl}
\big)
\cD^{(\bad)}_{kl}
~.
\label{3.13}
\eea
It  is obtained from the projection operator $\D^{(4)}_\rR$ defined 
in the curved-superspace case \cite{KLT-M11}
by switching off those torsion tensors which vanish in the AdS 
superspaces.\footnote{The fundamental property of $\D^{(4)}_\rR$ is that 
$Q_\rR^{(n)}:=\D_\rR^{(4)} T_\rR^{(n-4)}$  is a right weight-$n$ projective multiplet
for any right isotwistor superfield $T_\rR^{(n-4)} (v_\rR )$,
see \cite{KLT-M11} for more details.}

For any supergravity background, there is an alternative procedure \cite{KN}
 to construct a
composite right linear multiplet, $\bm G{}^{\bar{i}\bar{j}} $, from the left vector multiplet:
\bea
\bm G{}^{\bar{i}\bar{j}} 
= \frac{\ri}{4} (\cD^{i j \bar{i} \bar j} 
+ 8 \ri \cS^{ij\bar{i}\bar{j}} ) \Big( \frac{W_{ij}}{W_\rL}\Big) ~, \qquad 
W_\rL:= \sqrt{ W^{ij} W_{ij}}~.
\eea
It is applicable only in the case when $W_\rL$ is nowhere vanishing, 
$W_\rL \neq 0$. The superfield $\bm G{}^{\bar{i}\bar{j}} $ proves to be primary 
under the super-Weyl transformations \cite{KN}. Unlike $\bm G{}^{\bar{i}\bar{j}} $, 
our composite linear multiplet \eqref{3.8} exists only in the AdS superspaces. 
Its definition does not require $W^{ij}$ to be nowhere vanishing. 
These results are similar to those derived many years ago by Siegel 
for the 4D $\cN=2$ tensor multiplets
\cite{Siegel85}.


\section{SYM actions in $\cN=4$ AdS superspaces}
\setcounter{equation}{0}

To start with, we recall the locally supersymmetric action principle in $\cN=4$ 
matter-coupled supergravity \cite{KLT-M11}. 
In general,  the $\cN=4$ supersymmetric action may be presented as a sum of two terms, the 
left $S_\rL$ and right $S_\rR$ ones,
\bea
S= S_{\rm L} + S_{\rm R}~.
\eea 
The right action has the form 
\bea
S_{\rR}  &=& \frac{1}{2\pi} \oint_{\g_\rR} (v_\rR, \rd v_\rR)
\int \rd^3 x \,{\rm d}^8\q\,E\, C_\rR^{({-4})} \cL_\rR^{(2)}~, 
\qquad E^{-1}= {\rm Ber}(E_A{}^M)~,
\label{Action-right} 
\eea
where the Lagrangian $\cL_\rR^{(2)}(v_\rR)$ is a real right projective multiplet 
of weight  2.
The action involves a {\it model-independent} primary
 isotwistor superfield  $C_\rR^{(-4)} (v_\rR) $ 
defined to be real with respect to the smile-conjugation
and obey the differential equation\footnote{In conformal supergravity, 
the field $C_\rR^{(-4)} (v_\rR)$ has to be primary of weight $-2$ under the super-Weyl transformations 
\cite{KLT-M11}.}
\bea
\D_\rR^{(4)}C_\rR^{(-4)}=1~,
\label{D4C}
\eea
with $\D_\rR^{(4)}$ the covariant right projection operator. 
In AdS superspace, $\D_\rR^{(4)}$ is given by eq. \eqref{3.13}. 

To describe the dynamics of an Abelian left vector multiplet in a given $\cN=4$ 
AdS superspace, it suffices to make use of the right action only,  such that $S_\rL =0$. 
We choose 
\bea
\cL_\rR^{(2)}
= \hf V_\rR {\bm W}^{({2})}_\rR
=
-\frac{\ri } {24} V_\rR \cD^{(\bar 2)ij}W_{ij}
=
-\frac{\ri }{24}\cD^{(\bar 2)ij}\Big(V_\rR W_{ij}\Big)
~, 
\label{4.4}
\eea
where the composite right $\cO(2)$ multiplet is given by \eqref{3.8b}.
The action defined by eqs. \eqref{Action-right} and \eqref{4.4}
is manifestly invariant under all the isometries of the $\cN=4$ AdS superspace
under consideration. 

By applying the relations \eqref{WD4V} and \eqref{D4C},  
the action defined by eqs. \eqref{Action-right} and \eqref{4.4} may be rewritten in the form:
\bea
S  [V_\rR ] &=& 
\frac{1}{8\pi^2} \oint (v_\rR, \rd v_\rR)
\oint (\hat{v}_\rR, \rd \hat{v}_\rR)
\int \rd^3 x \,{\rm d}^8\q\,E\, 
 \frac{1}{(v_\rR,\hat{v}_\rR)^2}V_\rR(v_\rR)V_\rR(\hat{v}_\rR)
~.~~~~~~
\eea
This is similar to the action for the Abelian $\cN=2$ vector multiplet in
four dimensions constructed first 
in the rigid supersymmetric case  \cite{G-R} (see also \cite{K-double}) 
and later in supergravity \cite{K-08}.

The action defined by eqs. \eqref{Action-right} and \eqref{4.4} is valid 
for all the $\cN=4$ AdS superspaces. It turns out that alternative forms 
for the supersymmetric action exist in two special cases: 
(i) the (2,2) AdS superspace; and (ii) the critical (4,0) 
AdS superspace with $2S+X=0$. 

In the case of (2,2) AdS superspace, 
the theory can be described using a left action only,  such that $S_\rR =0$.
The left Lagrangian is 
\bea
\cL^{(2)}_\rL = \hf \frac{ W^{(2)}_\rL   W^{(2)}_\rL }{ d^{(2)}_\rL  }~,
\qquad d^{(2)}_\rL (v_\rL) := d^{ij} v_i v_j ~, \qquad 
\cD_\g^{k\bak} d^{ij} =0~,
\label{4.6}
\eea
for some background covariantly constant real symmetric spinor $d^{ij}$. 
The covariant constancy of $d^{ij}$ implies that $d^{\ij} \propto l^{ij}$, 
with $l^{ij}$ one of the parameters of  (2,2) AdS superspace, see
eq. \eqref{SSX22}. Without loss of generality,  $d^{\ij}$ and  $l^{ij}$
may be identified.
The Lagrangian \eqref{4.6} admits a trivial extension 
to the non-Abelian case:
\bea
\cL^{(2)}_\rL = \frac{1}{2d^{(2)}_\rL } \,
{\rm tr} \Big( W^{(2)}_\rL   W^{(2)}_\rL  \Big)~.
\label{4.7}
\eea

In the case of (4,0) AdS superspace with $2S+X=0$, 
the SU(2)$_\rL$ curvature vanishes, according to eqs. 
\eqref{N=4alg} and \eqref{SSX40},  and there exists a 
covariantly constant real symmetric spinor $d^{ij}$ such that
$\cD_\g^{k\bak} d^{ij} =0$.
As a result, in this case we  can again use Lagrangians \eqref{4.6} or \eqref{4.7}
to describe SYM theories. 

In conclusion, we comment on two different  schemes 
to extend our results to the non-Abelian case for any $\cN=4$ 
AdS superspace.  Similar to the 5D discussion in \cite{K2006}, 
a SYM action may be defined by its variation\footnote{To avoid cluttering 
of the equations, here we use notation $V$ for the right tropical prepotential $V_\rR$.}
\bea
\d S_{\rm SYM} [V ] &=& \frac{1}{2\pi} \oint  (v_\rR, \rd v_\rR)
\int \rd^3 x \,{\rm d}^8\q\,E\, C_\rR^{({-4})} \,
{\rm tr} \Big( \D V \cdot {\bm W}^{({\bar 2})}_+ \Big)~,
\label{4.8}
\eea
where we have defined
\bea
\D V: = \re^{- V } \d \re^{V}~, \qquad 
{\bm W}^{({\bar 2})}_+ := \re^{-\O_+} {\bm W}^{({\bar 2})} \re^{\O_+}~,
\qquad 
{\bm W}^{({\bar 2})}
=
-\frac{\ri}{12}{\frak D}^{(\bad)}_{ij}{\frak W}^{ij}~.
\label{4.9}
\eea 
Here  ${\frak W}^{ij}$ denotes the non-Abelian field strength,
and $ \d \re^{V}$ an arbitrary variation of the non-Abelian tropical prepotential.
For more details, including  the definition of $\O_+$, the reader should 
consult  appendix \ref{AppendixA}. The projective superfields
$\D V$ and ${\bm W}^{({\bar 2})}_+$ take their values in the Lie algebra of the gauge group 
and transform only under the $\l$-group as follows:
\bea
\D V' =  {\rm e}^{ {\rm i} \l  } \D V {\rm e}^{ {- \rm i} \l  }~, \qquad
({\bm W}^{({\bar 2})}_+ )' =  {\rm e}^{ {\rm i} \l  } {\bm W}^{({\bar 2})}_+ {\rm e}^{- {\rm i} \l  }  ~.
\label{4.10}
\eea
Now, making use of  \eqref{A.33} and 
the expression for the analyticity projection operator \eqref{3.13},
we obtain
\bea
{\bm W}_+^{({\bar 2})}
=
-\frac{\ri}{12}\cD^{(\bad)}_{ij}{\frak W}_+^{ij}
=
-\ri\D_\rR^{(\bar{4})}\Big( {\rm e}^{-\O_+}  \pa^{(-\bar 2)}   {\rm e}^{\O_+}  \Big)
~.
\eea
As a result,  the variation \eqref{4.8} can be rewritten in the form
\bea
\d S_{\rm SYM} [V ] &=& 
-\frac{\ri}{2\pi} \oint  (v_\rR, \rd v_\rR)
\int \rd^3 x \,{\rm d}^8\q\,E\, 
{\rm tr} \Big{[} \D V \, {\rm e}^{-\O_+}  \pa^{(-\bar 2)}   {\rm e}^{\O_+}   \Big{]}
~.
\eea

It may be shown that this variation \eqref{4.8} is integrable. 
The action $ S_{\rm SYM} [V ] $ is gauge invariant, since an infinitesimal  
gauge transformation \eqref{A.14} corresponds to the choice\footnote{The
transformations \eqref{4.10} and \eqref{4.13} are classical and quantum
realizations of the  gauge transformation within the background-quantum 
splitting, see e.g. \cite{GGRS}.}
\bea
\D V = \ri (\breve{\bm \l} - \bm \l ) ~, \qquad
\breve{\bm \l} := \re^{-V} \breve{\l}\re^V~, \qquad 
\bm \l := \l~, 
\label{4.13}
\eea 
for which the variation  \eqref{4.8} proves to vanish. 
However, we have not yet 
been able  to integrate it in a closed form in terms of $\cN=4$ superfields.  
An alternative approach to 
obtaining a closed-form expression for the SYM action is to make use of the superform construction, 
see \cite{KN} and references therein. 
 In appendix E, we explicitly integrate the variation \eqref{4.8} upon its reduction 
 to (2,0) AdS superspace.


\section{Reduction to (2,0) AdS superspace}\label{section5}
\setcounter{equation}{0}

Suppose there is a rigid supersymmetric field theory formulated in a given $\cN=4$ AdS superspace. 
As shown in \cite{BKT-M}, 
such a  dynamical system can always be reformulated 
as a supersymmetric theory realized in  $(2,0)$ AdS superspace, with 
two  supersymmetries hidden. 
In this section we give a brief review of the superspace reduction 
$\cN=4$ AdS $\longrightarrow$ (2,0) AdS, concentrating mainly on the 
decomposition of the $\cN=4$ AdS isometries into the (2,0) AdS isometries
and additional non-manifest symmetries. In the next section the reduction 
procedure will be applied to reformulate the $\cN=4$ theories in (2,0) AdS 
superspace.

We start by reminding the reader that the algebra of $\cN=4$ AdS covariant derivatives, eq. \eqref{N=4alg}, 
involves a covariantly constant tensor 
$\cS^{ij \bar i \bar j} = \cS^{(ij) (\bar i \bar j )}$. 
Its explicit form is 
given by \eqref{SSX--}. 
For all the $\cN=4$ AdS superspaces, it may be seen that
applying an $R$-symmetry transformation
allows us to choose several components of this tensor to vanish, 
\bea
\cS^{11\bau\bad}=\cS^{12\bau\bau}=\cS^{11\bau\bau}=\cS^{22\bad\bad}=0
~,
\label{condONcS}
\eea
as well as to have the property 
\bea
\cS+\cS^{12\bau\bad} = S~.
\label{condONcS2}
\eea
The proof of these claims was given in \cite{BKT-M}, 
and it will be reiterated  below. 
In this gauge,  the operators $\cD_a$,
$\cD_\a^{1\bau}$ and $(-\cD_\a^{2\bad})$
 form an algebra\footnote{Given 
a tensor superfield $U$
of Grassmann parity $\e(U)$, 
the operation of complex conjugation maps $\cD_\a^{1\bau} U$ to
$\overline{\cD_\a^{1\bau} U}=- (-1)^{\e(U) } \cD_{\a 1\bau}\bar U=- (-1)^{\e(U)}\cD_\a^{2\bad}\bar U$.}
which is isomorphic to that of (2,0) AdS superspace, eq.  
\eqref{20AdSsuperspace}, 
provided the U(1)$_R$ generator is identified with 
\bea
{ \cJ}:=\hat \cJ + \frac{X}{2S}\hat \cZ ~,
\label{mod-cJ}
\eea
where we have defined the operators 
\bea
\hat \cJ:=(\bL^{12}+\bR^{\bau\bad})~, \qquad
\hat \cZ:=(\bL^{12}-\bR^{\bau\bad})~, \qquad 
[\hat \cJ, \hat \cZ]=0
\eea
with the properties
\bsubeq
\bea
{[} \hat \cJ,\cD_\a^{1\bau}{]}&=&\cD_\a^{1\bau}
~,\qquad
{[} \hat \cJ,(-\cD_\a^{2\bad}){]}=-(-\cD_\a^{2\bad})~,
\\
{[}\hat \cZ,\cD_\a^{1\bau}{]}&=& 0~, \qquad ~~~
{[}\hat \cZ,(-\cD_\a^{2\bad}){]}=0
~.
\eea
\esubeq
The   generator $\cJ$ defined by (\ref{mod-cJ})
coincides with $\hat \cJ$ for all conformally flat $\cN=4$ AdS superspaces. 

Given an $\cN=4$ tensor superfield  $U(x,\q_{\imath\bar{\jmath}})$,
we define its projection to  (2,0) AdS superspace by
\bea
U|:=U(x,\q_{\imath \bar{\jmath}})|_{\q_{1\bad}=\q_{2\bau}=0}~.
\label{N2red-1}
\eea
By definition, $U|$  depends on the Grassmann coordinates
$\q^\mu:=\q^\mu_{1\bau}$ and their complex conjugates,  $\qb^\mu=\q^\mu_{2\bad}$.
We will refer to $U|$ as the bar-projection of  $U$. 
For the $\cN=4$ AdS covariant derivatives \eqref{N4-dev}
the bar-projection is defined as\footnote{Depending on the choice of parameters 
$\cS,\,\cS^{ij\bai\baj}$ and $X$,
the $R$-symmetry connection may take its values  
in a subgroup of SU(2)$_\rL\times$SU(2)$_\rR$. This point was discussed earlier. }
\bea
\cD_{{A}}|=E_{{A}}{}^{{M}}|\pa_{{M}}
+\hf\O_{{A}}{}^{bc}|\cM_{bc}
+\F_{{A}}{}^{kl}|\bL_{kl}
+\F_{{A}}{}^{\bak\bal}|\bR_{\bak\bal}~.
\label{N2red-3}
\eea
Since the algebra of operators $\big(\cD_a,\,\cD_\a^{1\bau},\,-\cD_\a^{2\bad}\big)$  is
isomorphic to that of the (2,0) AdS superspace, 
eq.~\eqref{20AdSsuperspace}, 
the freedom to perform general coordinate, local Lorentz and 
$R$-symmetry transformations may be used 
to choose a gauge in which
\bea
\cD_\a^{1\bau}|=\cD_\a
~,~~~~~~
-\cD_\a^{2\bad}|=\cDB_\a
~,
\label{7.24}
\eea
where 
$ \cD_\a$ and $ \bar \cD_\a$
are  the spinor covariant derivatives of  (2,0) AdS superspace \eqref{20derivatives}.

In the coordinate system defined by \eqref{7.24}, 
the operators $\cD_\a^{1\bau}|$ and $\cD_{\a}^{2\bad}|$  involve no
partial derivative with respect to $\q_{1\bad},\,\q_{2\bau}$.
Therefore, for any positive integer $k$,  
it holds that $\big( \cD_{\hat{\a}_1} \cdots  \cD_{\hat{\a}_k} U \big)\big|
= \cD_{\hat{\a}_1}| \cdots  \cD_{\hat{\a}_k}| U|$, 
where $ \cD_{\hat{\a}} :=\big( \cD_\a^{1\bau}, -{\cD}_\a^{2\bad} \big)$ 
and $U$ is a tensor superfield. This also implies that $\cD_a|$ 
coincides with the vector covariant derivative of (2,0) AdS superspace. 
The latter will be denoted by the same symbol $\cD_a$. 
We hope that no notational confusion will occur for the reader.

Let us fix an $\cN=4$ AdS superspace and consider its
Killing vector field $\x$ 
specified by  eqs. \eqref{sK-1} -- \eqref{sK-2}. 
We introduce the bar-projections of the parameters  involved: 
\bsubeq
\begin{gather}
\t^a:=\x^a|
~,~~~
\t^\a:=\x^\a_{\1\1}|~,~~~
\bar{\t}^\a=\x^\a_{\2\2}|~,~~~
t:=\ri(\L^{12}+\L^{\bau\bad})|=\overline{t}
~, ~~~t^{ab}:= \L^{ab}|~
;
\label{N2proj-20-1}
\\
\ve^\a:=-\x^\a_{1\bad}|~,~~~
\bar{\ve}^\a=\x^\a_{2\bau}|
~,~~~
\hat{\s}:= \ri (\L^{12}-\L^{\bau\bad})|=\bar{\hat{\s}}
~;
\label{N2proj-20-2-0}
\\
\bar{\ve}_\rL:=-\frac{1}{4S}\L^{11}|~,~~~
\ve_\rL=-\frac{1}{4S}\L^{22}|
~,~~~
\bar{\ve}_\rR=-\frac{1}{4S}\L^{\bau\bau}|~,~~~
\ve_\rR=-\frac{1}{4S}\L^{\bad\bad}|
~.~~~~~~
\label{N2proj-20-2}
\end{gather}
\esubeq
The parameters $(\t^a, \, \t^\a,\,\bar{\t}_\a,\, t ,\, t^{ab} )$
describe the infinitesimal isometries of  (2,0) AdS superspace. 
This may be proved by computing 
the bar-projection of the equations (\ref{sK-2-1})--(\ref{sK-2-4}).

The  parameters
$(\veps^\alpha, \,\bar\veps_\alpha, \,\hat{\sigma}, \,\veps_\rL, \,\bar\veps_\rL, \,
\veps_\rR, \,\bar\veps_\rR)$ generate those $\cN=4$ AdS isometries
which are not manifest in the (2,0) AdS setting. 
These include two rigid supersymmetries
and the residual $R$-symmetry transformations. 
Depending on the $\cN=4$ AdS superspace chosen,
these parameters obey different constraints.
Let us spell out these constraints
for the various cases.

\subsection{AdS superspace reduction (4,0) $\to$ (2,0)}
\label{4020}

For the reduction from (4,0) to (2,0) AdS superspace,
we find
a set of differential relations between $\veps_\alpha$, $\veps_\rL$, 
$\veps_\rR$ and their complex conjugates:
\bsubeq \label{extra4020}
\begin{gather}
\cD_\a\bar{\ve}_\b =
4S\ve_{\a\b}\,\bar{\ve}_\rL
~,~~~
\cDB_\a\ve_\b=
-4S\ve_{\a\b}\,{\ve}_\rL
~,
\label{extra4020-1}
\\
\cD_\a\ve_\b
=-4S\ve_{\a\b}\,\bar{\ve}_\rR
~,~~
\cDB_\a\bar{\ve}_\b
=4S\ve_{\a\b}\,{\ve}_\rR
~,~~~~~~
\label{extra4020-1b}
\\
\cDB_\a\ve_\rL =\cDB_\a{\ve}_\rR=0
~,~~~~~~
\cD_\a{\ve}_\rL=
\ri \,\ve_{\a}\Big(1+\frac{X}{2S}\Big)
~,~~~
\cD_\a{\ve}_\rR=-\ri\,\bar{\ve}_{\a}\Big(1-\frac{X}{2S}\Big)
~.~~~~~~
\label{extra4020-2}
\end{gather}
\esubeq
The action of the U$(1)_R$ generator  \eqref{mod-cJ}  on these parameters is
\bea
{ \cJ}\ve_\a=-\frac{X}{2S}\ve_\a~,\qquad
{ \cJ}{\ve}_\rL=-\Big(1+\frac{X}{2S}\Big){\ve}_\rL~,\qquad
{ \cJ}{\ve}_\rR=-\Big(1-\frac{X}{2S}\Big){\ve}_\rR~.
~~~~~~
\label{7.28}
\eea
The real parameter $\hat{\sigma}$, corresponding to one of the residual $R$-symmetries,
can be shown to obey 
\bea
\Big(\hat{\s}-\frac{X}{2S}t\Big) = \textrm{const}~.
\label{7.30}
\eea
A finite U(1)  transformation generated by the constant parameter $(\hat{\sigma} - t{X}/{2S} )$
does not act on the (2,0) AdS superspace.
It will be more convenient to parametrize this transformation 
using the constant parameter
\bea
\s:=\Big(\hat{\s}-\frac{X}{2S}t\Big)
~,
\eea
such that $t+\hat{\s}=\Big(1+\frac{X}{2S}\Big)t+\s$.

In the critical cases, $|X|=2S$, 
 the parameters are further constrained as follows:
\bsubeq
\bea
X=2S&:&~~~~~~
\L^{\bak\bal}=\ve_\rR=0~,~~~~~~
\cD_\a\ve_\b=
\cDB_\a\bar{\ve}_\b
=0~;
\\
X=-2S&:&~~~~~~
\L^{kl}=\ve_\rL=0~,~~~~~~
\cDB_\a\ve_\b=
\cD_\a\bar{\ve}_\b
=0~.
\eea
\esubeq

\subsection{AdS superspace reduction (3,1) $\to$ (2,0)}
\label{3120}

In order to carry out reduction from (3,1) to (2,0) AdS superspace,
a local $R$-symmetry transformation can be applied to choose
$w^{i\bai}$ of the form:
\bea
w^{1\bau}=w^{2\bad}=0~,~~~
w^{1\bad}=1~,~~~
w^{2\bau}=-\overline{(w^{1\bad})}=-1
~.
\eea
As a result,  the conditions \eqref{condONcS}  and \eqref{condONcS2} hold.
In the gauge chosen we have
\bea
&\L^{\bak\bal}=\d_{k}^{\bak}\d_{l}^{\bal}\L^{kl}
~,~~~
\ve_\rL=\ve_\rR:=\ve~.
\label{3120uuu}
\eea
Computing the bar-projection of (\ref{sK-2-1})--(\ref{sK-2-4}) gives
\bsubeq \label{extra3120}
\begin{gather}
\cD_\a\bar{\ve}_\b=-\cD_\a\ve_\b=
4S\ve_{\a\b}\,\bar{\ve}
~,~~~
\cDB_\a\ve_\b=-\cDB_\a\bar{\ve}_\b=
-4S\ve_{\a\b}\,\ve
~,
\label{extra3120-1}
\\
\cDB_\a\ve=0
~,~~~
\cD_\a\ve=
\frac{\ri}{2} \big(\ve_{\a}-\bar{\ve}_{\a}\big)
~.
\label{extra3120-2}
\end{gather}
\esubeq
These imply
\bea
\cD_\a(\ve_\b+\bar{\ve}_\b)=\cDB_\a(\ve_\b+\bar{\ve}_\b)=0
~.
\label{extra3120-1b}
\eea
The real parameter $\hat{\sigma}$ proves to vanish.

\subsection{AdS superspace reduction (2,2) $\to$ (2,0)}
\label{2220}

In order to carry out reduction from (2,2) to (2,0) AdS superspace,
a local $R$-symmetry transformation
can be applied to bring 
$l^{ij}$ and $r^{\bai\baj}$ to the form:
\bea
&l^{11}=l^{22}= 0~, ~~~r^{\bau\bau}=r^{\bad\bad}=0~,~~~
l^{12}=-\ri~,~~~r^{\bau\bad}=\ri
~.
\eea
As a result, the conditions \eqref{condONcS} and \eqref{condONcS2}  hold.
We then have 
\bea
&\ve_\rL=\L^{22}=0~,~~~
\ve_\rR=\L^{\bad\bad}=0~.
\label{2220uuu}
\eea
Computing the bar-projection of (\ref{sK-2-1}) -- (\ref{sK-2-4}) gives
\bea
&\cD_\a\ve_\b=\cDB_\a\ve_\b=0
~.
\label{extra2220-1}
\eea
The real parameter $\hat{\sigma}$ is constant.

\section{$\cN=4$ vector multiplet theories in $(2,0)$ AdS superspace}
\label{N4-20-Abelian}
\setcounter{equation}{0}

In this section we  reduce all results, which were obtained in sections 3 and 4 
within the manifestly  $\cN=4$ AdS supersymmetric setting, 
to (2,0) AdS superspace.

\subsection{The field strength}

We recall that the left vector multiplet is described by 
the gauged-invariant field strength $W^{ij}$, which is a left linear multiplet.
It can equivalently be described  by the left $\cO(2)$ multiplet 
$W^{(2)}(v_\rL):=v_iv_jW^{ij}$, with $v^i$ the homogeneous complex coordinates
for  ${\mathbb C}P^1$. It is useful 
to introduce an inhomogeneous complex coordinate 
$\z_\rL$ for ${\mathbb C}P^1$ by the rule 
\bea
\z_\rL:=\frac{v^2}{v^1}\in\mathbb{C}
~,~~~~~~
\eea
which is defined in the north chart of 
${\mathbb C}P^{1}$. 
Then we 
can represent 
the (2,0) AdS projection of  $W^{(2)}(v_\rL)$ as
\bea
W^{(2)}(v_\rL)|
&=&
\ri\z_\rL(v^{1})^2W^{[2]}(\z_\rL)|~, \qquad
W^{[2]}(\z_\rL)|
=
-\frac{\ri}{\z_\rL}\F
+G
-\ri\z_\rL\Fb
~,~~~~~~
\label{6.2}
\eea
where we have introduced the $\cN=2$ superfields
\bea
\F:=W^{22}|~,~~~~~~
G:=2\ri \,W^{12}|~,~~~~~~
\Fb=W^{11}|
~.
\eea
By projecting the analyticity constraint $\cD_\a^{(i\bai}W^{jk)}=0$ to (2,0) AdS superspace, it is not difficult to 
prove that $\F$ is chiral and  $G=\bar G$ is a real linear superfield, 
\bea
\cDB_\a\F=\cD_\a\Fb=0~,~~~~~~
\cD^2G=\cDB^2G=0
~.
\eea

The generators of the $R$-symmetry group ${\rm SU(2)}_\rL \times {\rm SU(2)}_\rR$
act on 
$W^{ij}$ by the rule
\bea
\bL^{ij}W^{kl}=\ve^{k(i}W^{j)l}+\ve^{l(i}W^{j)k}
~,~~~~~~
\bR^{\bai\baj}W^{kl}
=0
~.
\eea
Bar-projecting the first relation 
to (2,0) superspace gives
\bsubeq
\bea
&
\bL^{11}\F=\ri G~,~~~
\bL^{12}\F=-\F~,~~~
\bL^{22}\F=0~,
\\
&\bL^{11}G=-2\ri \Fb~,~~~
\bL^{12}G=0
~,~~~
\bL^{22}G=2\ri\F
~.
\eea
\esubeq
The fields 
$\F$ and $G$ are neutral under the right $R$-symmetry 
group  SU(2)$_\rR$.
This observation tells us that the U(1)$_R$ generator of (2,0) AdS superspace, 
eq. \eqref{mod-cJ}, acts on the superfields
introduced as follows:
\bea
\cJ\F=-{\bm q} \F~, \qquad
\cJ\Fb={\bm q} \Fb~, \qquad
\cJ G=0
~,
\eea
where we have defined
\bea
{\bm q}:=1+\frac{X}{2S}~.
\eea

Given an isometry transformation of the $\cN=4$ AdS superspace
(see appendix \ref{Isometries-4} for the details), the 
transformation laws of the field strength $W^{ij}$ is
\bea
\d_\cK W^{ij}
&=&
\cK W^{ij}
=\x^a\cD_a W^{ij}
+\x^\a_{k\bak}\cD_\a^{k\bak} W^{ij}
+2\L^{(i}{}_kW^{j)k}
~.~~~~~~
\eea
We now project this transformation law 
to (2,0) superspace.
Using the analyticity condition $\cD_\a^{(i\bai}W^{jk)}=0$
and the results of section \ref{section5}, we obtain
\bsubeq \label{6.10}
\bea
\d_\cK \F
&=&
\big(\t+\ri t\cJ\big) \F
+\ri\big(\ve^\a\cDB_\a -4S\ve_\rL\big) G
-\ri \s \F
~, \label{6.10a}
\\
\d_\cK G
&=&
\t G
-\ri\big(\bar{\ve}^\a\cD_\a
-8 S\bar{\ve}_\rL\big)\F
-\ri\big(\ve^\a\cDB_\a
-8 S\ve_\rL\big)\Fb \non \\
&=&
\t G
+\ri \cD_\a(\bar{\ve}^\a\F)
+\ri \cDB_\a(\ve^\a\Fb)
~.~~~~~~~~~
\eea
\esubeq
We recall that  the parameters $\t=\t^a\cD_a+\t^\a\cD_\a+\bar{\t}_\a\bar \cD^\a$ 
and $t$ describe the isometry of (2,0) AdS superspace. 
The relations \eqref{6.10}  are
 universal in the sense that they hold for  all the $\cN=4$ AdS superspaces.
All information about a concrete $\cN=4$ AdS superspace
is encoded 
in the Killing parameters $\ve_\rL$, $\ve_\a$ and $\s$,
which satisfy different constraints as described in the previous section.

\subsection{The tropical prepotential} 

The left field strength $W^{ij}$ is constructed in terms of the 
right weight-zero tropical  prepotential $V_\rR(v_\rR)$ according to 
eq.~\eqref{3.5}. We introduce an inhomogeneous complex coordinate 
$\z_\rR$ for   ${\mathbb C}P^1$ by the rule 
\bea
v^{\bai}=v^{\bau}(1,\z_\rR)~,\qquad
\z_\rR:=\frac{v^\bad}{v^\bau}
\in \mathbb{C}~.
\eea
We also choose the isospinor $u_\bai$ in \eqref{3.5} to be 
\bea
u_\bai=(1,0)
~.
\eea
Then the relation \eqref{3.5} becomes
\bea
W^{ij}&=&\frac{\ri}{4 }\oint \frac{\rd \z_\rR}{2\p}
\big(\cD^{i  j\bau\bau}-4\ri\cS^{ij\bau\bau}\big) V_\rR(\z_\rR)
~.
\label{prepot-2}
\eea
Here the right weight-zero tropical prepotential 
is described by 
the Laurent series
\bea
V_\rR(v_\rR)=
\sum_{k=-\infty}^{+\infty}(\z_\rR)^k V_k~,~~~~~~
\bar{V_k}=(-1)^k V_{-k}~.
\label{6.14}
\eea
The analyticity constraint \eqref{A.11}
projected to (2,0) AdS implies 
\bea
\cD_\a^{1\bad}V(\z_\rR)|=\z_\rR\cD_\a V(\z_\rR)|
~,~~~~~~
\cD_\a^{2\bau}V(\z_\rR)|=-\frac{1}{\z_\rR}\cDB_\a V(\z_\rR)|
~.
\label{6.15}
\eea

The $\cN=4$ AdS transformation law of the tropical prepotential \cite{KLT-M11} is 
\bea
\d_\cK V_\rR
= \Big( \x^a\cD_a 
+\x^\a_{k\bak}\cD_\a^{k\bak}  +  \L^{\bar i \bar j} \bR_{\bar i \bar j} \Big) V_\rR~,
\label{6.16}
\eea
where 
\bea
\L^{\bar i \bar j} \bR_{\bar i \bar j}  V_\rR = - \L^{(\bar 2)} \pa^{(-\bar 2)}V_\rR~, 
\qquad
 \L^{(\bar 2)} = \L^{\bar i \bar j}v_{\bar i} v_{\bar j}~,
 \eea
 and the operator $\pa^{(-\bar 2)}$ is defined according to  \eqref{A.26}.
Projecting the transformation law \eqref{6.16} to (2,0) AdS superspace gives
 \bea
\d_\cK V_\rR(\z_\rR)|
&=&
\big(\t+\ri t \cJ\big) V_\rR(\z_\rR)|
-\z_\rR\ve^{\a} \cD_\a V_\rR(\z_\rR)|
+\frac{1}{\z_\rR}\bar{\ve}_{\a} \cDB^\a V_\rR(\z_\rR)|
\non\\
&&
+\ri\Big{[}
4\ri S\ve_\rR\, \frac{1}{\z_\rR}
-\s
 +4\ri   S\bar{\ve}_\rR \z_\rR
\Big{]}
 \,\z_\rR\pa_{\z_\rR} V_\rR(\z_\rR)|
 ~.~~~~~~
\label{6.18}
\eea
Note that the action of $\cJ$  on $V(\z_\rR)$ is
\bea
\cJ V_\rR(\z_\rR)
&=&
(2-\bm q )
\bR^{\bau\bad} V_\rR(\z_\rR)
=(2-\bm q )\z_\rR\pa_{\z_\rR} V_\rR(\z_\rR)
~.
\eea

For the coefficients in 
the Laurent series expansion of $V_\rR|$, eq. \eqref{6.14},
the transformation law \eqref{6.18} leads to  
\bea
\d_\cK V_k|
&=&
\big(\t+\ri t \cJ\big) V_k|
-\ri\s k V_k|
\non\\
&&
-\Big(\ve^{\a} \cD_\a  +4 (k-1)  S\bar{\ve}_\rR \Big) V_{k-1}|
+\Big(\bar{\ve}_{\a} \cDB^\a-4(k+1) S\ve_\rR \Big)V_{k+1}|
~.~~~~~~
\label{6.20}
\eea

Evaluating the contour integral in \eqref{prepot-2} and making use of the analyticity 
condition \eqref{6.15}, 
it is possible to obtain the expression for $\F$, $\Fb$ and $G$ in terms of $V_k|$.
The results are:
\bea
\F=\frac{1}{4}\cDB^2 V_{1}|
~,~~~
\Fb=-\frac{1}{4}\cD^2 V_{-1}=\frac{1}{4}\cD^2 \bar{V}_{1}|
~,~~~~~~
G=
\frac{\ri}{2 }
\cD^\a\cDB_\a V_0|
~.
\label{6.21}
\eea
These relations show 
that the components of the gauge-invariant field strength are constructed
in terms of only three components of 
the tropical prepotential: 
$V_1$, $\bar{V}_{1}$ and $V_0$.
It is easy to see that the other components of the tropical prepotential, 
$V_2$, $V_3, \dots$, are purely gauge degrees of freedom.

Let us first  compute the isometry transformation of $V_1|$ 
by applying  \eqref{6.20}:
\bea
\d_\cK V_1|
&=&\,
\big(\t+\ri t \cJ\big) V_1|
-\ve^{\a} \cD_\a  V_{0}|
+\big(\bar{\ve}_{\a} \cDB^\a-8 S\ve_\rR \big)V_{2}|
-\ri\s V_1|
~.
\eea
This is equivalent to
\bea
\d_\cK V_1|
&=&
\big(\t+\ri t \cJ\big) V_1|
-\ve^{\a} \cD_\a  V_{0}|
-\ri\s V_1|
+\cDB_\a\big(\bar{\ve}^{\a}  V_{2}|\big)
~.
\label{6.28} 
\eea
The last term is a pure gauge transformation that does not contribute to 
$\F=\frac{1}{4}\cDB^2 V_{1}|$. From eq. \eqref{6.28}
we deduce
\bea
\d_\cK \F
&=&
\big(\t+\ri t \cJ\big) \F
-\frac{1}{4}\cDB^2(\ve^{\a} \cD_\a  V_{0}|)
-\ri\s\F
~.
\eea
This may be seen to be equivalent to \eqref{6.10a}.

Next we compute the isometry transformation of $V_0|$ 
using  \eqref{6.20}
\bea
\d_\cK V_0|
&=&
\t V_0|
+\big(\bar{\ve}_{\a} \cDB^\a-4 S\ve_\rR \big)V_{1}|
+\big(\ve^{\a} \cD_\a  -4S\bar{\ve}_\rR \big) \bar{V}_{1}|
~.
\eea
This can be rewritten, with the aid of the identities 
\bea
\ve_\rR=-\frac{1}{8S}\cDB_\a\bar{\ve}^\a
~,~~~~~~
\bar{\ve}_\rR=-\frac{1}{8S}\cD^\a\ve_\a~,
\eea
as follows
\bea
\d_\cK V_0|
&=&
\t V_0|
+\Big{\{}
\bar{\ve}_{\a}\cDB^\a V_{1}|
+\hf(\cDB_\a\bar{\ve}^{\a}) V_{1}|
+{\rm c.c.}
\Big{\}}
 ~. ~~~~~~~~~
\label{6.24}
\eea
This transformation law is valid 
for all the $\cN=4$ AdS superspaces.

\subsubsection{(4,0) AdS supersymmetry}
\label{621}

In the case of  (4,0) AdS supersymmetry with  
$\bm q \neq 0$, the following relations hold:
\bea
\ve_\a=-\frac{\ri}{{\bm q}}\cD_\a\ve_\rL
~,~~~
\bar{\ve}_\a=\frac{\ri}{{\bm q}}\cDB_\a\bar{\ve}_\rL
~.
\eea
Then the transformation of $V_0$, eq. \eqref{6.24}, 
can be rewritten as 
\bea
\d_\cK V_0|
=
\t V_0|
-\frac{2\ri}{{\bm q}}\big(\bar{\ve}_\rL \F-\ve_\rL \Fb\big)
+\frac{\ri}{2{\bm q}}\cDB^2(\bar{\ve}_\rL V_{1}|)
-\frac{\ri}{2{\bm q}}\cD^2(\ve_\rL \bar{V}_{1}|)
~.  ~~~~~~~~~
\eea
The last two terms 
generate a pure gauge transformation and 
can be omitted.

In the case of critical (4,0) AdS supersymmetry,
$X+2S = 2S{\bm q}=0$, we have $\ve_\rL=0$.
Here $\ve_\a$ can be expressed in terms of a real parameter $\r$ such that
\bea
\ve_{\a}=-\ri\cD_\a\r~,~~~
\cD^2\r
=8\ri S\bar{\ve}_\rR~, ~~~ \bar \r = \r
~.
\eea
The parameter $\r$ is defined up to an arbitrary constant shift of the form 
\bea
\r ~\to ~\r +  \j ~,  \qquad \cJ \j =0 ~, 
\qquad 
\overline{\j}=\j={\rm const}
~.
\label{6.311}
\eea
Using $\r$, we can rewrite 
the transformation of $V_0|$, eq. \eqref{6.24}, 
as 
\bea
\d_\cK V_0|
&=&
\t V_0|
-2\ri\r\big(\F-\Fb\big)
+ \hf \Big{\{}
{\ri}(\cDB^2\r V_{1}|)
+{\rm c.c.}
\Big{\}}
~.
\eea
The last term generates a pure gauge transformation of $V_0|$, 
and so does the shift \eqref{6.311}.

\subsubsection{(3,1) AdS supersymmetry}

In the case of (3,1) AdS supersymmetry, 
we can introduce a complex parameter $\r$ such that
\bea
\cD_\a\r=
\frac{\ri}{2} \big(\ve_{\a}+\bar{\ve}_{\a}\big)~,\qquad \cJ \r = -\r~, \qquad 
\cD^2\r=0~.
\label{6.27} 
\eea
The existence of this representation follows from eqs. 
\eqref{extra3120-1} and \eqref{extra3120-1b}.
The parameter $\r$ is defined modulo arbitrary shifts of the form
\bea
\r ~\to ~\r + \bar \j ~,  \qquad \cJ \bar \j = - \bar \j ~, 
\qquad \cD_\a \bar \j =0~.
\label{6.31}
\eea
Due to eqs. \eqref{extra3120-2} and \eqref{6.27}, 
the spinor parameter  $\ve_\a$ can now be expressed in the form
\bea
\ve_\a=-\ri\cD_\a(\ve+\r)=-\ri\cDB_\a(\bar{\ve}-\bar{\r})
~,~~~~~~\bar{\ve}_\a=\ri\cDB_\a(\bar{\ve}+\bar{\r})=\ri\cD_\a(\ve-\r)
~.
\eea
Using this representation allows us to rewrite the transformation law
\eqref{6.24} as
\bea
\d_\cK V_0|
&=&
\t V_0|
-2\ri(\bar{\ve}+\bar{\r})\F +2\ri (\ve +\r ) \bar \F 
\non \\
&&
+ \Big\{\frac{\ri}{2}\cDB^2\Big((\bar{\ve}+\bar{\r}) V_{1} | \Big)
+{\rm c.c.}
\Big{\}}
 ~. ~~~~~~~~~
\label{6.33}
\eea
The expression in the second line generates a pure gauge transformation and 
can be omitted. 
It should be pointed out that any shift of $\r$ defined by \eqref{6.31} leads to 
a pure gauge transformation of $V_0|$. 

\subsubsection{(2,2) AdS supersymmetry}

It remains to consider the case of  the (2,2) AdS supersymmetry. 
In accordance with \eqref{extra2220-1}, 
we can introduce a complex parameter ${\r}$ such that
\bea
\ve_{\a}&=&-\ri\cD_\a {\r}
~,\qquad \cJ \r = -\r~, \qquad
\cD^2 {\r}
=0~.
\label{6.37}
\eea
As in the (3,1) case, this parameter is defined modulo arbitrary antichiral shifts 
of the form \eqref{6.31}.
Then the transformation law \eqref{6.24} can be rewritten as
\bea
\d_\cK V_0|
&=&
\t V_0|
-2\ri (\bar{{\r}}\,\F -\r \bar \F) + 
\Big\{ \frac{\ri}{2}\cDB^2\big(\bar{\r}  V_{1}| \big)
+{\rm c.c.}
\Big{\}}
 ~. ~~~~~~~~~
\eea
Here the third term generates a pure gauge transformation and 
can be omitted.

\subsection{The composite right linear multiplet}

One of the main aims of the present section is 
to reduce the action for $\cN=4$ SYM to (2,0) AdS.
It involves 
the composite right $\cO(2)$ multiplet ${\bm W}_\rR^{(2)}$, which is defined by 
\eqref{3.8b} and can be represented as
\bea
{\bm W}^{(2)}_\rR
=
\ri\z_\rR(v^\bau)^2{\bm W}^{[\bar 2]}
~,~~~~~~
{\bm W}^{[\bar 2]}(\z_\rR)=
-\frac{\ri}{2\z_\rR}{\bm W}^{\bad\bad}
+2\ri{\bm W}^{\bau\bad}
-\ri\z_\rR{\bm W}^{\bau\bau}
~.~~~~~~
\label{6.399}
\eea
Computing the bar-projection of the superfields on the right  gives
\bsubeq
\bea
{\bm W}^{\bau\bau}|
&=&
-\frac{\ri}{4}\cD^2 \F
+\cS^{22}{}^{\bau \bau }\Fb
~,~~~~~~
{\bm W}^{\bad\bad}|
=
\frac{\ri}{4}\cDB^2\Fb
+\cS^{11}{}^{\bad \bad }\F
~,
\\
{\bm W}^{\bau\bad}|
&=&
-\frac{1}{4}\Big{(}
\cD^{\a}\cDB_\a
+4\ri \bm q\cS
\Big{)} G
~.
\eea
\esubeq
The values of $\cS^{22}{}^{\bau \bau }$ and $\cS^{11}{}^{\bad \bad }$
corresponding to the various types of $\cN=4$ AdS supersymmetry are:
\bsubeq
\bea
&\text{(4,0) AdS:}&~~~\cS^{22\bau\bau}=\cS^{11\bad\bad}=0
~;
\\
&\text{(3,1) AdS:}&~~~\cS^{22\bau\bau}=\cS^{11\bad\bad}=-S
~;
\\
&\text{(2,2) AdS:}&~~~\cS^{22\bau\bau}=\cS^{11\bad\bad}=0
~.
\eea
\esubeq

\subsection{The $\cN=4$ vector multiplet actions}

It was proven in \cite{BKT-M} that the reduction of the right  action, eq.~\eqref{Action-right}, 
to (2,0) AdS superspace is given by 
\bea
S_\rR &=& \int \rd^3x\, \rd^2{\q} \rd^2{\qb}\, {  E}\, 
\oint_C \frac{\rd \zeta_\rR}{2\pi \ri \zeta_\rR}\, \cL^{[2]}_\rR(\z_\rR)|
~,~~~~~~
{ E}^{-1}:={\rm Ber}( {E}_A{}^M)~,
\label{AdS-N4-2_0_action}
\eea
where $\cL_\rR^{[2]}$ is related to the original Lagrangian by the rule
$\cL_\rR^{(2)}(v_\rR) =\ri\z_\rR(v^\bau)^2\cL_\rR^{[2]} (\z_\rR)$.
In the $\cN=4 $ SYM case, the Lagrangian is given by \eqref{4.4}.
Its reduction to (2,0) AdS superspace is
\bea
\cL_\rR^{[2]}|
=\hf \Big(V(\z_\rR){\bm W}_\rR^{[2]}(\z_\rR)\Big)\big|
~.
\label{6.40}
\eea

It is now a simple exercise to compute the contour integral 
in the action defined by \eqref{AdS-N4-2_0_action} and \eqref{6.40}.
We obtain
\bea
S = \int \rd^3x\, \rd^2\q  \rd^2\qb\, E &\Big{[}&
\frac{1}{8}V_1\Big(\cDB^2\Fb
-\,4\ri\cS^{11}{}^{\bad \bad }|\F
\Big)
-\frac{1}{4}V_0\big{(}
\ri\cD^{\a}\cDB_\a
- 4\cS \bm q
\big{)} G
\non\\
&&
+\frac{1}{8}\bar{V}_{1}\Big(\cD^2\F
+\,4\ri\cS^{22}{}^{\bau \bau }|\Fb
\Big)
\Big{]}
~.
\eea
This is equivalent to
\bea
S &=& 
\int \rd^3x\, \rd^2\q  \rd^2\qb\, E \, \Big{[}
\Fb\F
-\frac{1}{2}G^2
+ \cS \bm qV_0G
\Big{]}
\non\\
&&
+\frac{\ri}{2}\int \rd^3x\, \rd^2\q  \, { \cE} \, \cS^{11}{}^{\bad \bad }\F^2
-\frac{\ri}{2}\int \rd^3x\, \rd^2\qb\, {\bar{{ \cE}}}\,\cS^{22}{}^{\bau \bau }\Fb^2
~.
\label{4gen}
\eea
We see that the action involves a Chern-Simons term for $\bm q \neq 0$ and 
a mass-like chiral term for $\cS^{11}{}^{\bad \bad } \neq 0$.


In the (2,2) AdS case,  the SYM theory can equivalently be described by the left Lagrangian \eqref{4.7}. Let us 
prove this claim in the Abelian case by 
comparing 
the two different actions upon their reduction to (2,0) AdS superspace. 
Upon this reduction, 
the action associated with \eqref{4.7} becomes
\bea
S^{(2,2)} &=& 
 \int \rd^3x\, \rd^2{\q} \rd^2{\qb}\, {  E}\, 
\oint_C \frac{\rd \zeta_\rL}{2\pi \ri \zeta_\rL}\, 
\frac{1}{2d^{[2]}_\rL } 
\big(W^{[2]}_\rL   W^{[2]}_\rL\big) \big|
~.
\label{22W2}
\eea
Here $W^{[2]}_\rL(\z_\rL)|$ is given by eq. \eqref{6.2}. 
We remind the reader that $d^{ij} $ is proportional to $l^{ij}$ 
and the latter has the only non-zero component  $l^{12}=-\ri$. 
By choosing 
\bea
d^{ij}=-\hf l^{ij}
~,~~~~~~
d^{[2]}=-1
~,
\eea
it is trivial to compute the contour integral in \eqref{22W2}. The resulting action is
\bea
S^{(2,2)} &=&
\int \rd^3x \rd^2{\q} \rd^2{\qb}\, {  E}\, 
 \,
\Big{[}
\Fb\F
-\hf G^2
\Big{]}
~.
\eea
This coincides with  the action \eqref{4gen} in the (2,2) AdS case.


\section{Concluding comments}
\setcounter{equation}{0}

In this paper we have constructed the pure $\cN=4$ SYM theories in three dimensions 
for all types of $\cN=4$ AdS supersymmetry.\footnote{A brief discussion of 
off-shell $\cN=4$ hypermultiplets coupled to the vector multiplet 
is given in appendix A.} 
In the Abelian case, these theories were described within the manifestly $\cN=4$ 
supersymmetric setting as well as in (2,0) AdS superspace where only $\cN=2$
supersymmetry is manifest. In all the $\cN=4$ AdS superspaces, 
the vector multiplet action has 
the universal form given by eqs. \eqref{Action-right} and \eqref{4.4}. 
This is an example of the right linear multiplet action involving a special 
composite linear multiplet, eq. \eqref{3.8b}.\footnote{The left and right 
linear multiplet actions \cite{KLT-M11} are known to be universal 
in the sense that the action of any off-shell $\cN=4$ supergravity-matter system
may be realized as a sum of left and right linear multiplet actions \cite{KN}.}
All specific details of the theory are encoded in the type of $\cN=4$ AdS 
supersymmetry chosen. 
These differences become explicit when  the theory is reformulated in (2,0) AdS 
superspace in which the $\cN=4$ vector multiplet
decomposes into 
the $\cN=2$  vector multiplet
described by a real linear superfield $G$
and the  chiral scalar $\F$ and its conjugate $\bar \F$. 
The latter multiplets are equivalently described by  
unconstrained gauge prepotentials $\cX:=V_1|$ and $\cV:=V_0| = \bar \cV$
such that $\F=\frac{1}{4}\cDB^2 \cX$ and $G=\frac{\ri}{2}\cD^\a\cDB_\a \cV$.
Let us now summarize the key properties of the theory for all types 
of $\cN=4$ AdS supersymmetry.

In the case of  (4,0) AdS supersymmetry with  $\bm q = 1 +X/2S \neq 0$, 
the non-manifest supersymmetry transformations  are\footnote{In the critical case with $X=2S$ and 
${\bm q}=2$, the last expression in \eqref{40tF} is not defined. 
In this case $\d_\ve \F$
can be represented as
$\d_\ve \F=\frac{1}{2}\cDB^2(\o G)$.
Here the real parameter $\o$ is such that 
$\ve_{\a}=-\ri\cDB_\a\o$
and
$\cDB^2\o=-8\ri S\ve_\rL$.}
\bsubeq \label{7.1}
\bea
\d_\ve \F
&=&
\ri\big(\ve^\a\cDB_\a -4S\ve_\rL\big) G
=-\frac{1}{4}\cDB^2(\ve^{\a} \cD_\a  \cV)
=-\frac{1}{2(2-{\bm q})}\cDB^2(\bar{\ve}_\rR G)
~,
\label{40tF}
\\
\d_\ve \cV
&=&
-\frac{2\ri}{{\bm q}}( \bar{\ve}_\rL \F - \ve_\rL \bar \F) 
\label{40tV}
~, 
\eea
and therefore
\bea
\d_\ve G
&=&
-\ri\big(\bar{\ve}^\a\cD_\a
-8 S\bar{\ve}_\rL\big)\F
+{\rm c.c.}
=
\ri \cD_\a(\bar{\ve}^\a\F)
+{\rm c.c.}
=\frac{\ri}{2}\cD^\a\cDB_\a\,\d_\ve \cV~.
\label{40tG}
\eea
\esubeq
The parameters 
 $\ve_\rL $ and $\ve_\a$ 
  are defined in section \ref{4020}.
The invariant action is
\bea
S^{(4,0)} &=& 
\int \rd^3x\, \rd^2\q  \rd^2\qb\, E \, \Big{[}
\Fb\F
-\frac{1}{2}G^2
+ S \bm q \cV G
\Big{]}~.
\label{7.2}
\eea
This is exactly the AdS analogue of the Abelian $\cN=4$ SYM theory on $S^3$
recently constructed in \cite{SS}. The Chern-Simons term in \eqref{7.2} 
is generated due to the non-zero curvature of AdS$_3$. 
It disappears in the flat-superspace limit. 
Thus our results explain 
the mysterious structure of $\cN=4$ supersymmetric Yang-Mills theories on $S^3$ 
discovered in \cite{SS}.

In the case of critical (4,0) AdS supersymmetry with
$X+2S= 2S{\bm q}=0$, we have  $\ve_\rL=0$ and 
the non-manifest supersymmetry transformations are
\bsubeq \label{7.3} 
\bea
\d_\ve \F
&=&
\ri\ve^\a\cDB_\a G
=-\frac{1}{4}\cDB^2(\ve^{\a} \cD_\a  \cV)
~,
\\
\d_\ve \cV
&=&
-2\ri\r\big(\F-\Fb\big)
~,
\eea
and therefore
\bea
\d_\ve G
&=&
-\ri\bar{\ve}^\a\cD_\a\F
+{\rm c.c.}
=
\ri \cD_\a(\bar{\ve}^\a\F)
+{\rm c.c.}
=\frac{\ri}{2}\cD^\a\cDB_\a\d_\ve\cV
~.
\eea
\esubeq
The action is given by  \eqref{7.2} 
with ${\bm q}=0$.

In the case of (3,1) AdS supersymmetry, the 
non-manifest supersymmetry transformations  
are 
\bsubeq \label{7.4}
\bea
\d_\ve \F
&=&
\ri\big(\ve^\a\cDB_\a -4S\ve\big) G
=-\frac{1}{4}\cDB^2(\ve^{\a} \cD_\a  \cV )
=-\frac{1}{2}\cDB^2\big(
(\bar{\ve}-\bar{\r})G
\big)
~,
\\
\d_\ve \cV
&=&
-2\ri(\bar{\ve}+\bar{\r})\F +2\ri({\ve}+{\r}) \bar \F 
~,
\eea
and therefore
\bea
\d_\ve G
&=&
-\ri\big(\bar{\ve}^\a\cD_\a
-8 S\bar{\ve}\big)\F
+{\rm c.c.}
=
\ri \cD_\a(\bar{\ve}^\a\F)
+{\rm c.c.}
=\frac{\ri}{2}\cD^\a\cDB_\a\,\d_\ve \cV~.
\eea
\esubeq
The invariant action is
\bea
S^{(3,1)} &=& 
\int \rd^3x\, \rd^2\q  \rd^2\qb\, E \, \Big{[}
\Fb\F
-\frac{1}{2}G^2
+\hf S\cV G
\Big{]}
- \hf {S}\Big{\{} {\ri}  \int \rd^3x\, \rd^2\q \,  { \cE}\, \F^2
+{\rm c.c.}~\Big{\}}
~.~~~~~~~~~
\label{7.5}
\eea
This theory  possesses a chiral mass-like term, which is 
a new feature as compared with the $\cN=4$ SYM on $S^3$ \cite{SS}.

In the case of (2,2) AdS supersymmetry, 
the non-manifest  supersymmetry transformations are\footnote{In 
 this case we do not need to consider 
the transformation of $\cV$ since it does not appear in the action.}
\bsubeq \label{7.6}
\bea
\d_\ve \F
&=&
\ri\ve^\a\cDB_\a G
=\frac{1}{2}\cDB^2\big(
\bar{\r}G
\big)
~,
\\
\d_\ve G
&=&
-\ri\bar{\ve}^\a\cD_\a\F
+{\rm c.c.}
=\cD^\a\cDB_\a\big(
\bar{\r}\,\F-\r\, \bar \F 
\big)
~.
\eea
\esubeq
Here the parameter $\r$ is defined by \eqref{6.37}.
The invariant action is
\bea
S^{(2,2)} &=& 
\int \rd^3x\, \rd^2\q  \rd^2\qb\, E \, \Big{[}
\Fb\F
-\frac{1}{2}G^2
\Big{]}
~.
\eea
No Chern-Simons term shows up because the theory is formulated entirely 
in term of the field strength $W^{ij}$ in $\cN=4$ AdS superspace, see eq. 
\eqref{4.7}. 

The chiral mass-like term appears only in the (3,1) AdS case. 
In all other cases it is prohibited by the rigid U(1) symmetry 
\bea
\d \F = - \ri \s \F ~.
\eea

In the non-Abelian case, we defined the $\cN=4$ SYM action by its variation, 
eqs. \eqref{4.8} and \eqref{4.9}, induced by an arbitrary variation of the tropical 
prepotential.\footnote{We constructed the manifestly $\cN=4$ SYM actions 
 in the cases of (2,2) and critical (4,0) AdS supersymmetries.} 
So far we have not yet been able 
to integrate this variation in a closed form in terms of $\cN=4$ superfields.  
 However, there are three  obvious ways to obtain closed-form 
 expressions for the $\cN=4$ SYM action for all types of $\cN=4$ AdS supersymmetry.
 Firstly, it may be achieved using the superform construction in complete analogy with 
 the Chern-Simons results of \cite{KN}. 
Secondly, we may start with the $\cN=4$ tropical prepotential $V_\rR (v_\rR)$ 
and the gauge covariant field strength ${\frak W}^{ij}$ and then reduce 
their $\cN=4$ isometry transformations to (2,0) AdS superspace. 
Using the explicit structure of the non-manifest supersymmetry transformations 
derived, there is a standard procedure to reconstruct a closed-form expression for 
the $\cN=4$ SYM action in $\cN=2$ superspace.\footnote{Unlike \cite{SS}, 
here we do not have to guess the structure of 
two non-manifest supersymmetry transformations, 
we derive them from first principles.} This procedure is 
explicitly implemented in appendix 
D in which we present complete $\cN=4$ SYM actions 
in (2,0) AdS superspace for all types of $\cN=4$ AdS supersymmetry. 
Thirdly, we may reduce the variation of the SYM action, eqs. \eqref{4.8} and 
\eqref{4.9}, to (2,0) AdS superspace, in which the variation may be readily 
integrated. This is explicitly done in appendix E. 

In this paper we have focused our attention on the $\cN=4$ left vector multiplet.
Analogous  results for the right vector multiplet can be  obtained by applying the 
 mirror map \cite{Zupnik99,KLT-M11}.

In this paper we have studied the $\cN=4$ SYM theories.
In \cite{KLT-M12,BKT-M} 
the off-shell formalism, which was developed for general 3D $\cN=3$ supergravity-matter systems  
\cite{KLT-M11},  was applied to the (3,0) and (2,1) AdS cases. 
Using these techniques allows one to  construct  $\cN=3$ SYM 
theories for both types of $\cN=3$ AdS supersymmetry,
along the same lines as in the present paper. 
\\

\noindent
{\bf Acknowledgements:}\\
SMK is grateful to Igor Samsonov and Dima Sorokin for showing him 
a preliminary draft of their paper \cite{SS} and for asking questions that 
stimulated the research presented in this note. 
We thank Joseph Novak for reading the manuscript. 
The work of SMK was supported in part by the ARC Discovery
projects DP1096372 and DP140103925. 
The work of GT-M was  supported by the Australian 
Research Council's Discovery Early Career 
Award (DECRA) No. DE120101498 and by 
the ARC Discovery project DP140103925.

\appendix 

\section{$\cN=4$ SYM  and projective superspace}
\label{AppendixA}

In this appendix we consider a left $\cN=4$ Yang-Mills supermultiplet 
in a conformal supergravity background \cite{KLT-M11} 
and uncover the origin of a tropical prepotential $V_\rR (v_\rR)$. 
Our consideration is similar to that given by Lindstr\"om and Ro\v{c}ek in the case 
of 4D $\cN=2$ SYM theory \cite{LR2}. Only right projective supermultiplets appear in 
this section. For this reason we consistently avoid using a subscript `R'  and 
simply denote  $V_\rR$ by $V$ etc.

\subsection{Tropical prepotential}

To describe a left Yang-Mills supermultiplet, we 
introduce  gauge covariant derivatives
\bea 
{\frak D}_A = \cD_A +\ri \,{\frak A}_A ~,
\eea
where $\cD_A$ denotes  the $\cN=4$ supergravity covariant derivatives \cite{KLT-M11}, and 
the  connection ${\frak A}_A(z) $ takes values in the Lie algebra
of the gauge group. 
The fact that we are dealing with the left vector multiplet, is encoded 
in the anti-commutation relation: 
\bea
\{{\frak D}_\a^{i\bai}, {\frak D}_\b^{j\baj }\}&=& \dots
+2\ve_{\a\b} \ve^{\bai \baj } \frak{W}^{ij}
~,
\eea
where the ellipsis denotes the purely supergravity terms. 
The SYM field strength $\frak{W}^{ij} = \frak{W}^{ji}$ is Hermitian, 
$(\frak{W}^{ij})^\dagger = \frak{W}_{ij}$, 
and obeys the Bianchi identity 
\bea
{\frak D}^{(i \bar i}_\g \frak{W}^{jk)} =0~.
\eea
Under the gauge group (to be referred to as the $\t$-group),
the covariant derivatives and any covariant matter superfield multiplet $U (z)$
transform as follows
\bea
{\frak D}'_A={\rm e}^{{\rm i}\tau} {\frak D}_A e^{-{\rm i}\tau} ~,
\qquad U'={\rm e}^{{\rm i}\tau}U~, \qquad \tau=\t^\dagger ~, 
\label{A.4}
\eea
with the Lie-algebra-valued gauge parameters $\t (z) $ being Hermitian
and otherwise  unconstrained. In particular, the field strength transforms as 
\bea
\frak{W}^{ij}{}'  = {\rm e}^{{\rm i}\tau} \frak{W}^{ij} {\rm e}^{-{\rm i}\tau}~.
\eea 

Using an isospinor $v:=v^\bai \in {\mathbb C}^2 \setminus  \{0\}$, 
which provides homogeneous  coordinates for ${\mathbb C}P^1$, 
we introduce gauge covariant operators 
\bea
{\frak D}_\a^{(\bar 1) i}:=v_{\bar i} {\frak D}_\a^{i\bai}~,
\label{A.6}
\eea
in complete analogy with \eqref{3.2}. It is easy to see that the anti-commutator 
$\{{\frak D}_\a^{(\bar 1)i} , {\frak D}_\b^{(\bar 1)j} \} $ coincides with 
the right-hand side of  \eqref{3.9a}, i.e. it does not involve the gauge field. 
This means that we may represent ${\frak D}_\a^{(\bar 1) i}$ in the form:
\bea
{\frak D}_\a^{(\bar 1) i} = \re^{\O_+} \cD_\a^{(\bar 1) i} \re^{-\O_+}~, 
\label{A.7}
\eea
where we have introduced a Lie-algebra-valued bridge superfield
\bea
\O_{+}(\z)  = \sum_{n=0}^{\infty} \O_n \z^n~, \qquad \z:=\frac{v^{\bar 2}}{v^{\bar 1}}~.
\label{A.8}
\eea
Another representation for ${\frak D}_\a^{(\bar 1) i} $ follows 
by applying the smile-conjugation to \eqref{A.7} (see, e.g., \cite{KLT-M11} for the definition 
of the smile-conjugation). The result is 
\bea
{\frak D}_\a^{(\bar 1) i} = \re^{-\O_-} \cD_\a^{(\bar 1) i} \re^{\O_-}~, \qquad 
\O_{-} (\z) 
= \sum_{n=0}^{\infty} (-1)^n \, \O_n^\dagger \,
\frac{1}{ \z^n }~.
\label{A.9}
\eea
Introduce a Lie-algebra-valued superfield $V(\z)$ defined by 
\bea
{\rm e}^V := {\rm e}^{\O_-} {\rm e}^{\O_+} ~, \qquad 
V(\z ) 
=\sum_{n=-\infty}^{\infty} V_n \z^n~, \qquad V_n^\dagger = (-1)^n V_{-n}~.
\eea
It may be seen from \eqref{A.7} and \eqref{A.9} that $V$ is a covariant projective multiplet, 
\bea
\cD_\a^{(\bar 1) i} V = 0~. \label{A.11}
\eea

It follows from \eqref{A.4} and \eqref{A.7} that 
 the gauge transformation law of $\O_+$ is 
\bea
{\rm e}^{\O_+' (\z)} =  {\rm e}^{{\rm i}\t }  {\rm e}^{\O_+(\z) }  {\rm e}^{ -{\rm i} \l (\z) } ~,
\eea
where the new gauge parameter $\l(\z) $ is a covariant weight-zero arctic multiplet
\bea
\cD_\a^{(\bar 1) i} { \l} = 0~, 
\qquad 
{ \l}(\z) =  \sum_{n=0}^{\infty} {\l}_n \z^n~.
\eea
The gauge transformation law of the tropical prepotential is 
\bea
{\rm e}^{V' } =  {\rm e}^{{\rm i}\breve{\l} }  {\rm e}^{V }  {\rm e}^{ -{\rm i} \l  } ~.
\label{A.14} 
\eea
We see that $V$ transforms under the $\l$-group only. 

\subsection{Polar hypermultiplets} 

$\cN=4$ supersymmetric matter may be described in terms of gauge-covariantly
 arctic multiplets and their smile-conjugate antarctic multiplets. 

A gauge-covariantly arctic multiplet of weight $n$, ${\bm \U}^{(\bar n)} (v)$,
is defined by 
\bea
{\frak D}^{(\bar 1) i }_{ \a} {\bm \U}^{(\bar n)} =  0~, 
\qquad 
{\bm \U}^{(\bar n )}(v) = (v^{\bar 1})^n  \sum_{k=0}^{\infty} {\bm \U}_k \z^k~.
\eea
It can be represented in the form
\bea
{\bm \U}^{(n)} (v )= {\rm e}^{\O_+(\z)} \U^{(n)} (v)~, 
\eea
where $\U^{(n)} (v) $ is an ordinary covariant arctic multiplet of weight $n$ 
(see \cite{KLT-M11} for more details), 
\bea
{\cD}^{(\bar 1) i}_{ \a}  \U^{(\bar n)}   =0~, 
\qquad 
{ \U}^{(\bar n )} (v) =   (v^{\bar 1})^n\sum_{k=0}^{\infty} {\U}_k \z^k~.
\eea

A gauge-covariantly antarctic multiplet of weight $n$, $\breve{\bm \U}{}^{(\bar n)}(v) $, 
is defined by 
\bea
\breve{\bm \U}^{(\bar n)}  \stackrel{\longleftarrow}{{\frak D}^{(\bar 1) i}_{ \a}} =0~, 
\qquad 
\breve{\bm \U}{}^{(\bar n)}(v) =  (v^{\bar 2})^n \sum_{k=0}^{\infty} (-1)^k{\bm \U}_k^\dagger \frac{1}{\z^k}~.
\eea
It can be represented in the form
\bea
\breve{\bm \U}^{(\bar n)} (v )=\breve{\U}^{(\bar n)}(v)  {\rm e}^{\O_-(\z)}   ~, 
\eea
where $\breve{\U}^{(\bar n)} (v) $ is an ordinary antarctic multiplet 
\bea
{\cD}^{(\bar 1) i}_{ \a}\breve{ \U}^{(\bar n)}  =0~, 
\qquad 
\breve{ \U}^{(\bar n)}(v) =  (v^{\bar 2})^n \sum_{n=0}^{\infty} (-1)^n{ \U}_n^\dagger \frac{1}{\z^n}~.
\eea

The gauge-covariantly arctic multiplet of weight $n$, ${\bm \U}^{(\bar n)}(v) $, and its smile-conjugate antarctic 
one, 
 $\breve{\bm \U}^{(\bar n)} (v) $, constitute the gauge-covariantly polar multiplet of weight $n$. 
The gauge transformation laws of  ${\bm \U}^{(\bar n)} (v) $ and  $\breve{\bm \U}{}^{(\bar n)} (v) $
are 
\bea 
 {\bm \U}^{(\bar n)}{}' (v)= \re^{\ri \t} {\bm \U}^{(\bar n)}(v)~, 
 \qquad \breve{\bm \U}{}^{(\bar n)}{}' (v)= \breve{\bm \U}{}^{(\bar n)}  (v) \re^{-\ri \t } ~.
\eea 
The gauge transformation laws of  ${ \U}^{(\bar n)} (v) $ and  $\breve{ \U}^{(\bar n)} (v) $
are 
\bea
{ \U}^{(\bar n)}(v) = \re^{\ri \l (\z)} { \U}^{(\bar n)}(v) ~, \qquad
\breve{ \U}{}^{(\bar n)}{}' (v)= \breve{ \U}{}^{(\bar n)}    (v)\re^{-\ri \l(\z) }~.
\eea
In the case of weight $n=1$, a gauge invariant hypermultiplet Lagrangian
can be constructed. It is
\bea
\cL^{(\bar 2)} = \ri \breve{\bm  \U}{}^{(\bar 1)} {\bm \U}^{(\bar 1)} 
= \ri \breve{ \U}{}^{(\bar 1)}  \re^V { \U}^{(\bar 1)} ~.
\eea

\subsection{Arctic and antarctic representations}

Here we show that the SYM gauge connection ${\frak A}_A$ may be expressed in terms
of the tropical prepotential $V (\z)$, modulo the $\t$-gauge freedom. 
Our analysis in this subsection is inspired by the famous paper by Zupnik 
\cite{Zupnik87}.

Let us introduce a new isospinor $ u_{\bar i} \in {\mathbb C}^2 \setminus \{0\}$,
which is only required to obey the inequality $(v,u):= v{}^{\bar i}u_{\bar i} \neq 0$. 
Since $v^{\bar i}$ and $u^{\bar i}$ 
are linearly independent vectors, 
we can construct a new basis for the gauge covariant spinor derivatives
that includes ${\frak D}^{(\bar 1) i }_{ \a}$ and the following operators:
\bea 
{\frak D}^{(-\bar 1) i}_{ \a} := \frac{1}{(v,u)}u_{\bar i} \,{ \frak D}^{i \bar i}_{ \a} ~.
 \eea
It can be seen that 
\bea
\{ {\frak D}^{(\bar 1) i }_{ \a} , {\frak D}^{(-\bar 1) j }_{ \b} \}
=  \dots
-2\ve_{\a\b} \frak{W}^{ij} ~, 
\label{A25}
\eea
where the ellipsis denotes the purely supergravity terms. 

We introduce the first-order differential operators
\bea
\pa^{(\bar 2)} :=(v,u) \, v^{\bar i} \frac{\pa}{\pa u^{\bar i}} ~, \qquad
\pa^{(-\bar 2)} := \frac{1}{(v,u)} u^{\bar i} \frac{\pa}{\pa v^{\bar i}} 
\label{A.26}
\eea
such that 
\bea
[ \pa^{(\bar 2)} , \pa^{(- \bar 2)} ] = v^{\bar i} \frac{\pa}{\pa v^{\bar i}} 
- u^{\bar i} \frac{\pa}{\pa u^{\bar i}}  \equiv \pa^{(\bar 0)}~.
\eea
These operators are invariant under the $\t$-group.
It is easy to see that 
\bea
[\pa^{(-\bar 2)} , {\frak D}^{( \bar 1) i}_{ \a} ] = {\frak D}^{(-\bar 1) i }_{ \a}~. 
\label{A.28}
\eea

When dealing with polar hypermultiplets, it is useful to introduce 
an arctic representation  defined by the transformation  
\bea
\hat \cO \to \hat{\cO}_+ := {\rm e}^{-\O_+} \hat{\cO}\,{\rm e}^{\O_+} ~, 
\qquad U \to U_+ := {\rm e}^{-\O_+} U
\label{A.29}
\eea
applied to any operator $\hat \cO$ and covariant superfield $U$.
In the arctic representation, any gauge-covariantly arctic multiplet 
${\bm \U}^{(\bar n)} (v)$
becomes the ordinary arctic one, ${\U}^{(\bar n)} (v)$, 
\bea
{\bm \U}^{(\bar n)} (v) \to { \U}^{(\bar n)} (v)~, \qquad
\breve{\bm \U}^{(\bar n)} (v ) \to \breve{\U}^{(\bar n)}(v)  {\rm e}^{V(\z)}   ~.
\eea
The gauge covariant derivatives ${\frak D}^{(\bar 1) i }_{ \a} $ turn into 
the AdS  spinor covariant derivatives,
\bea
{\frak D}^{(\bar 1) i}_{ \a} \to {\cD}^{(\bar 1) i}_{ \a} ~. 
\eea
The important point is that the projective derivative $\pa^{(-\bar 2)} $ turns into
the operator 
\bea
\pa^{(-\bar 2)} \to {\frak D}^{(-\bar 2)} := \pa^{(-\bar 2)} 
+  {\rm e}^{-\O_+} ( \pa^{(-\bar 2)}   {\rm e}^{\O_+} )~,
\eea
which transforms as a covariant derivative under the $\l$-group. 
Then making use of  \eqref{A25}
in conjunction with 
$[{\frak D}^{(-\bar 2)} , {\frak D}^{( \bar 1) i}_{ \a} ] = {\frak D}^{(-\bar 1) i }_{ \a}$,
we read off
\bea
\frak{W}_+^{ij} = \frac{1}{4} \Big( {\cD}^{(\bar 2) ij} 
-4\ri\cS^{(\bad)}{}^{ij}
\Big)
\Big( {\rm e}^{-\O_+}  \pa^{(-\bar 2)}   {\rm e}^{\O_+}  \Big)~.
\label{A.33}
\eea
It may be seen that $\frak{W}_+^{ij} $ is independent of $u^{\bar i}$, 
$\pa^{(\bar 2)} \frak{W}_+^{ij} =0$. It also satisfies the property 
\bea
 {\frak D}^{(-\bar 2)} \frak{W}_+^{ij} =0~,
 \eea
 since in the original representation $\frak{W}^{ij}$ is independent of $v^{\bar i}$.
The field strength  obeys the Bianchi identity 
\bea
{\frak D}^{\a (i \bar i}_+ \frak{W}_+^{jk)} =0~.
\eea

If the gauge group is Abelian, then  
$\frak{W}^{ij}=\frak{W}_+^{ij}$ 
and \eqref{A.33} turns into
\bea
\frak{W}^{ij} 
= 
\frac{1}{4}  
\Big( {\cD}^{(\bar 2) ij} 
-4\ri\cS^{(\bad)}{}^{ij}
\Big)
\pa^{(-\bar 2)}
\O_+
~.
\eea
Since $\O_+$ is a homogeneous function of $v_\rR $ of degree zero, 
we have $\O_+ (v_\rR ) = \O_+ (\z) $ and
\bea
\pa^{(-\bar 2)}\O_+(v_\rR)
=
-\frac{1}{(v^\bau)^2}
\pa_\z\O_+(\z)
~.
\eea
Taking into account 
the fact that $\frak{W}^{ij}$ is independent of $\z$, 
we end up with 
the expression
\bea
\frak{W}^{ij} 
=
-\frac{1}{4} \Big( 
{\cD}^{ij\bad\bad}
-4\ri\cS^{ij}{}^{\bad\bad}
\Big)
\O_1
=
\frac{1}{4 }\big(\cD^{i  j\bau\bau}-4\ri\cS^{ij\bau\bau}\big)\O_{-1}
~.
\eea
This expression may be shown 
to be equivalent to \eqref{prepot-2}.

In complete analogy with the arctic representation, eq. (\ref{A.29}), 
one can introduce the antarctic representation defined by 
\bea
\hat \cO \to \hat{\cO}_- := {\rm e}^{\O_-} \hat{\cO}\,{\rm e}^{-\O_-} ~, 
\qquad U \to U_- := {\rm e}^{\O_-} U~.
\eea
In this representation, the SYM field strength takes the form
\bea
\frak{W}_-^{ij} = \frac{1}{4}\Big( 
 {\cD}^{(\bar 2) ij} 
-4\ri\cS^{(\bad)}{}^{ij}
\Big)
\Big( {\rm e}^{\O_-}  \pa^{(-\bar 2)}   {\rm e}^{-\O_-}  \Big)~.
\eea
Comparing the above with \eqref{A.33} gives 
\bea
\frak{W}_-^{ij} = {\rm e}^{V} \frak{W}_+^{ij} {\rm e}^{-V}~.
\eea


\section{Isometries of $\cN=4$ AdS superspaces} 
\setcounter{equation}{0}
\label{Isometries-4}

In this appendix we review the structure of the Killing vector 
fields of a given $\cN=4$ AdS superspace following \cite{BKT-M}. 

Given a particular  $\cN=4$ AdS superspace,
its isometry group  is generated by Killing vector 
fields, $\x=\x^a\cD_a+\x^\a_{i\bai}\cD_\a^{i\bai}$, obeying the Killing equation
\bea
0&=&
{[}
\cK,
\cD_A
{]}
~,~~~~~~
\cK:=\x
+\hf \L^{\g\d}\cM_{\g\d}
+\L^{kl}\bL_{kl}
+\L^{\bak\bal}\bR_{\bak\bal}
~.
\label{sK-1}
\eea
This  equation is equivalent to
\bsubeq
\bea
\cD_\a^{ i\bai}\x_{\b\g}&=&
4\ri\ve_{\a(\b}\x_{\g)}^{i\bai}
~,
\label{sK-2-1}
\\
\cD_\a^{i\bai}\x_{\b}^{j\baj}&=&
\x_{\a\b}\Big(\ve^{ij}\ve^{\bai\baj}\cS+\cS^{ij}{}^{\bai\baj}\Big)
+\hf\L_{\a\b}\ve^{ij}\ve^{\bai\baj}
+\L^{ij}\ve^{\bai\baj}\ve_{\a\b}
+\L^{\bai\baj}\ve^{ij}\ve_{\a\b}
~,
\label{sK-2-2}
\\
\cD_\a^{i\bai} \L_{\b\g}&=&
8\ri\ve_{\a(\b}\x_{\g) j\baj}(\cS^{ij}{}^{\bai \baj }+\ve^{ij}\ve^{\bai \baj }\cS)
~,
\label{sK-2-1b}
\\
\cD_\a^{i\bai}\L^{kl}&=&
-2\ri\ve^{i(k}\x_{\a}{}^{l)\bai}(2\cS+X)
-2\ri\x_{\a}{}^{i}{}_{\baj}\cS^{kl}{}^{\bai \baj}
~,
\label{sK-2-3}
\\
\cD_\a^{i\bai}\L^{\bak\bal}&=&
-2\ri\ve^{\bai(\bak}\x_{\a}{}^{i\bal)}(2\cS-X)
-2\ri\x_{\a j}{}^{\bai}\cS^{ij}{}^{\bak\bal}
~,
\label{sK-2-4}
\eea
\esubeq
and
\bsubeq \label{sK-2}
\bea
\cD_a\x_b&=&
\L_{ab}
~,
\label{sK-2-5}
\\
\cD_{a}\x^\b_{j\baj}
&=&
-\big(\cS\x^\g_{j\baj}
+\cS_{jk\baj\bak}\x^{\g k\bak}\big)(\g_a)_\g{}^\b
~,
\label{sK-2-6}
\\
\cD_a\L^{bc}
&=&
4S^2\big(\d_a^b\x^{c}
-\d_a^c\x^{b}
\big)
~,
\label{sK-2-7}
\\
\cD_a\L^{kl}&=&\cD_a\L^{\bak\bal}=0
~.
\label{sK-2-8}
\eea
\esubeq
Some useful implications of the above equations are
\bsubeq
\bea
&
\cD_{(\a}^{i\bai}\x_{\b\g)}
=\cD_{(\a}^{i\bai} \L_{\b\g)}=0
~,
\label{sK-2-1-b}
\\
&\cD^{\b i\bai}\x_{\a\b}=
6\ri\x_{\a}^{i\bai}~,~~~~~~
\cD^{\b i\bai} \L_{\a\b}=
12\ri\x_{\a j\baj}(\cS^{ij}{}^{\bai \baj }+\ve^{ij}\ve^{\bai \baj }\cS)
~,
\label{sK-2-1b-b}
\\
&\cD_{(\a}^{(i\bai}\x_{\b)}^{j)}{}_{\bai}=
\cD_{(\a}^{i(\bai}\x_{\b)}{}_i{}^{\baj)}=
0
~,~~~
\cD_{(\a}^{(i(\bai}\x_{\b)}^{j)\baj)}
=\x_{\a\b}\cS^{ij}{}^{\bai\baj}~,~~~
\cD_{(\a}^{i\bai}\x_{\b)}{}_{i\bai}=
4\x_{\a\b}\cS
+2\L_{\a\b}
~,~~~
\label{sK-2-5-ba}
\\
&\cD^{\a i\bai}\x_{\a i\bai}=\cD^{\a (i(\bai}\x_{\a}^{ j)\baj)}=0
~,~~~
\cD^{\a (i\bai}\x_{\a}^{j)}{}_{\bai}=
-4\L^{ij}
~,~~~
\cD^{\a i(\bai}\x_{\a}{}_i{}^{\baj)}=
-4\L^{\bai\baj}
~.
\label{sK-2-5-bb}
\eea
\esubeq
Here we have written the results in a form valid for the (4,0), (3,1) and (2,2) cases.
Depending on the $\cN=4$ AdS superspace under consideration, 
$\cS,\,X$ and $\cS^{ij\bai\baj}$ are constrained by 
(\ref{SSX40})--(\ref{SSX22}),
while the SU(2)$_\rL$ and SU(2)$_\rR$ parameters $\L^{kl}$ and $\L^{\bak\bal}$
are restricted by
\bsubeq
 \bea
&& (4,0)~~{\rm with}~~X=\phantom{+}2S:~~~\L^{\bak\bal}=0~;~~~~~~\\
&& (4,0)~~{\rm  with}~~X=-2S:~~~\L^{kl}=0~;
 ~~~~~~
 \\
&& (3,1):~~~\L^{\bak\bal}=w_k{}^{\bak}w_l{}^{\bal}\L^{kl}
 ~;
 \label{7.14c0}
 \\
&& (2,2):~~~
\L^{kl}=l^{kl}\L_\rL~,~~\overline{(\L_\rL)}=\L_\rL~,~~~~~~
\L^{\bak\bal}=r^{\bak\bal}\L_\rR~,~~\overline{(\L_\rR)}=\L_\rR~.
\label{7.14c}
 \eea
\esubeq


\section{Geometry of (2,0) AdS superspace}
\setcounter{equation}{0}

In this appendix we collect the main results concerning the (2,0) AdS superspace 
following \cite{KT-M11,BKT-M}.

The geometry of (2,0) AdS superspace is encoded in its  covariant derivatives
\bea
\cD_A = (\cD_a , \cD_\a, \bar \cD^\a)= E _A{}^M \pa_M +\hf {\O}_A{}^{cd}\cM_{cd}
+ \ri \,{ \F}_A \cJ
\label{20derivatives}
\eea 
obeying the following (anti-)commutation relations: 
\bsubeq \label{20AdSsuperspace}
\bea
& \{\cD_\a,\cD_\b\} =  0 ~, \qquad \{\bar \cD_\a,\bar \cD_\b\}
=0
~, \label{C.2a}
\\
&
\{\cD_\a,\bar \cD_\b\}
=
-2\ri \cD_{\a\b}
-4\ri \,S\, \ve_{\a\b} \cJ
+4\ri \,S\, \cM_{\a\b} ~, ~~~
\label{C.2b}
\\
& {[}\cD_{a},\cD_\b{]}
=
\,S\, (\g_a)_\b{}^\g\cD_{\g}~, 
\qquad
 {[}\cD_{a},\bar \cD_\b{]}
=
\,S\, (\g_a)_\b{}^\g\bar \cD_{\g}~,
\label{AdS_(2,0)_algebra_2}  \\
& {[}\cD_a,\cD_b{]} = -4\, S^2\, \cM_{ab}~.
\label{AdS_(2,0)_algebra_3} 
\eea
\esubeq
The generator $\cJ$ in \eqref{20AdSsuperspace}
corresponds to the gauged $R$-symmetry group, ${\rm U}(1)_R$, 
and acts on the covariant derivatives as
\bea
  {[}\cJ,\cD_\a{]}=\cD_\a~,\qquad
    {[}\cJ,\bar \cD _\a{]}=-\bar \cD_\a   ~.
 \eea

The isometries of the (2,0) AdS superspace are described  
by Killing vector fields, 
$\t=\t^a\cD_a+\t^\a\cD_\a+\bar{\t}_\a\bar \cD^\a$, obeying the  equation
\bea
\Big{[}\t+\ri t\cJ+\hf t^{bc}\cM_{bc},\cD_A\Big{]}=0~,
\label{2_0-Killing_iso-def}
\eea
for some parameters $t$ and $t^{ab}$. Choosing  $\cD_A=\cD_a$ in
\eqref{2_0-Killing_iso-def} gives
\begin{subequations}  \label{Killing_2.6}
\bea
\cD_a t &=& 0~, \label{Killing_2.6a}\\
\cD_a \t_b&=&   t_{ab}  ~, \label{Killing_2.6b} \\
 \cD_a \t^\b &=& - S \t^\g (\g_a )_\g{}^\b~, \label{Killing_2.6c} \\
\cD_a t^{bc} &=& 4S^2 (\d_a{}^b \t^c - \d_a{}^c \t^b) ~. \label{Killing_2.6d}
\eea
\end{subequations}
Eq. \eqref{Killing_2.6b} implies the standard Killing equation 
\bea
\cD_{a} \t_{b } +\cD_b \t_a=0~,
\eea
while \eqref{Killing_2.6c} is a Killing spinor equation. From  \eqref{Killing_2.6b}
and \eqref{Killing_2.6d} it follows that 
\bea
\cD_a \cD_b \t_c = 4S^2 (\eta_{ab} \t_c - \eta_{ac}\t_b)~.
\eea
Next, choosing $\cD_A=\cD_\a$ in
\eqref{2_0-Killing_iso-def} gives
\begin{subequations}\label{Killing_2.8}
\bea
\cD_\a \bar \t_\b &=& 0~, \label{Killing_2.8a}\\
\cD_\a t &=&  4S\bar \t_\a  ~, \label{Killing_2.8b} \\
 \cD_\a t^{\b \g}&=& - 4\ri S (\d_\a{}^\b \bar \t^\g + \d_\a{}^\g \bar \t^\b  ) ~, \label{Killing_2.8c} \\
\cD_\a \t^{\b \g} &=& -2\ri  (\d_\a{}^\b \bar \t^\g + \d_\a{}^\g \bar \t^\b) ~, \label{Killing_2.8d}\\
\cD_\a  \t^\b &=& \hf t_\a{}^\b +S \t_\a{}^\b +\ri \d_\a{}^\b t ~. \label{Killing_2.8e}
\eea
\end{subequations}
These equations have a number of nontrivial implications including the following:
 \bsubeq \label{2_0-Killing_iso}
\bea
&
\cD_{(\a}\t_{\b\g)}=\cD_{(\a} t_{\b\g)}=0~, 
\label{2_0-Killing_iso_2} \\
&\cD_{(\a}\t_{\b)}=-\bar \cD_{(\a}\bar{\t}_{\b)}=\dfrac{1}{2} t_{\a\b}+ S \t_{\a\b} ~,  
\label{2_0-Killing_iso_4} \\
&  \t_\a=
\dfrac{\ri}{6} 
\bar \cD^\b\t_{\a\b}=\dfrac{\ri}{12S}\bar \cD^\b t_{\a\b}~,
\label{2_0-Killing_iso_1}
\\
&
\cD_\g\t^\g
=
-\bar \cD^\g\bar{\t}_\g
=2\ri t ~.
\label{2_0-Killing_iso_3}
\eea
\esubeq
It follows from the above equations that the Killing superfields $\t^\a$, $t$ and $t^{ab}$ 
are given in terms of the vector parameter $\t^a$. Its components defined by $\t^a|_{\q=0}$ and 
$(-\cD^b \t^a) |_{\q=0}$ describe the isometries of AdS$_3$. The other isometry transformations 
of the (2,0) AdS superspace are contained not only in $\t^a$ but also in, e.g., 
the real scalar $t$ subject to the following equations:
\bea
\cD^2 t = \bar \cD^2 t =0~, \qquad (\ri \cD^\a \bar \cD_\a -8S) t =0~, \qquad
\cD_a t=0~.
\eea
At the component level, $t$ contains the  real constant parameter $t |_{\q=0}$ 
and the complex Killing spinor $\cD_\a t|_{\q=0}$,
which generate the $R$-symmetry  and supersymmetry transformations of  
the (2,0) AdS superspace respectively.


\section{$\cN=4$ SYM theories in (2,0) AdS superspace}
\setcounter{equation}{0}

In  this appendix we provide complete $\cN=4$ SYM actions in (2,0) AdS superspace
for all types of $\cN=4$ AdS supersymmetry. 
These actions are natural extensions of the $\cN=4$ vector multiplet models
derived in section \ref{N4-20-Abelian}.
 We start by recalling the structure of the $\cN=2$ Yang-Mills supermultiplet
 \cite{HKLR,ZP} as formulated in (2,0) AdS superspace. 
 
\subsection{$\cN=2$ SYM multiplet}
 
To describe a Yang-Mills supermultiplet in (2,0) AdS superspace,  
we introduce gauge covariant derivatives\footnote{We 
use one and the same symbol,  $\frak{D}_A$, 
to denote $\cN=2$ and $\cN=4$ gauge covariant derivatives, 
the latter have been introduced in appendix \ref{AppendixA}.
We hope no confusion may occur, since only the $\cN=2$ operators are used 
in the present appendix.}
\bea 
{\frak D}_A = \cD_A +\ri \,{\frak A}_A ~,
\eea
where $\cD_A$ stands for  the covariant derivatives of (2,0) AdS 
superspace, and 
the gauge connection ${\frak A}_A(z) $ takes values in the Lie algebra
of the gauge group. 
The anti-commutators of two spinor gauge covariant derivatives are constrained \cite{HKLR,ZP} by 
\begin{subequations}
\bea
&\{{\frak D}_\a, {\frak D}_\b \}=0~, \qquad 
\{\bar{\frak D}_\a, \bar{\frak D}_\b \} = 0~, \label{D.2a} \\
&\{{\frak D}_\a, \bar{\frak D}_\b \}=\cdots
+\ri \ve_{\a\b}\frak{G}
~,~~~~~~
\label{BianchiG}
\eea
\end{subequations}
where the ellipsis denotes the right-hand side of \eqref{C.2b}.  
The SYM field strength $\frak{G}$ is Hermitian, $\frak{G}^\dagger = \frak{G}$, 
and obeys the Bianchi identity
\bea
0={\frak D}^2 \frak{G}=\bar{\frak D}^2\frak{G}
~.
\eea
The gauge group  is defined to act on
the covariant derivatives and any  matter multiplet $U (z)$
as follows
\bea
{\frak D}'_A={\rm e}^{{\rm i}\tau} {\frak D}_A \re^{-{\rm i}\tau} ~,
\qquad U'={\rm e}^{{\rm i}\tau}U~, \qquad \tau=\t^\dagger ~, 
\eea
where the Lie-algebra-valued gauge parameters $\t (z) $ is  only constrained
to be Hermitian. The field strength transforms in the adjoint representation, 
\bea
\frak{G}' = {\rm e}^{{\rm i}\tau} \frak{G} {\rm e}^{-{\rm i}\tau} ~.
\eea
The gauge group will be referred to as  the $\t$-group.

The constraints \eqref{D.2a} are solved in complete analogy with the 
4D $\cN=1$ case (see, e.g., \cite{GGRS}) as follows:
\bea
\bar {\frak D}_\a = \re^\O \bar \cD_\a \re^{-\O}~, \qquad
 {\frak D}_\a = \re^{-\O^\dagger}  \cD_\a \re^{\O^\dagger}~.
 \label{D.6}
\eea
Here $\O (z) $ is an  unconstrained complex Lie-algebra-valued 
bridge superfield. Its gauge freedom is larger than the $\t$-group:
\bea
\re^{\O'} =  {\rm e}^{{\rm i}\tau} \re^\O {\rm e}^{-\ri \l}~, \qquad
\bar \cD_\a \l =0~.
\eea
Under the $\l$-transformation introduced, the gauge covariant derivatives
\eqref{D.6} remain unchanged. 

Let $\bm \F$ be a gauge-covariantly chiral 
scalar superfield, $\bar {\frak D}_\a \bm \F =0$, transforming in the adjoint 
representation of the gauge group. It may be represented in the form
\bea
{\bm \F} = \re^\O \F \re^{-\O} ~, \qquad \bar \cD_\a \F =0~.
\eea
Here the chiral scalar $\F$ is independent of the gauge field. 
It is inert under the $\t$-transformations and changes under the $\l$-transformations
by the rule
\bea
\F' = \re^{\ri \l} \F \re^{-\ri \l}~.
\eea
The Hermitian conjugate of $\bm \F$ is a gauge-covariantly antichiral  superfield
$\bar {\bm \F}$ constrained by  ${\frak D}_\a \bar {\bm \F}=0$.
Its explicit form is 
\bea
\bar {\bm \F} :=   {\bm \F}^\dagger = \re^{-\O^\dagger}  \F^\dagger \re^{\O^\dagger}~,
\qquad \cD_\a \F^\dagger =0~.
\eea 

It is often advantageous to use a chiral representation defined by the transformation 
\bea
\hat{\cO} \to \re^{-\O} \hat{\cO} \re^{\O} ~, 
\qquad U \to \re^{-\O} U~, 
\label{D11}
\eea
which has to be applied to any operator $\hat{\cO} $ and covariant superfield $U$. 
In this representation, the gauge covariant spinor derivatives look like 
\bea
{\frak D}_\a=\re^{-\cV}\cD_\a\re^{\cV}
~, \qquad
\bar{\frak D}_\a=\cDB_\a~, 
\label{D12}
\eea
and 
the adjoint multiplets $\bm \F$ and $\bar {\bm \F}$ turn into
\bea
\bm \F = \F~, \qquad \bar {\bm \F} = \re^{-\cV} \F^\dagger \re^\cV ~.
\eea
Here we have introduced the Hermitian Lie-algebra-valued prepotential  $\cV$   defined by 
\bea
\re^\cV := \re^{\O^\dagger}\re^\O~, \qquad \cV ^\dagger = \cV~.
\label{D14}
\eea
The virtue of the chiral representation is that the gauge field is described
in terms of a single prepotential, $\cV$, with the gauge transformation law
\bea
\re^{\cV'} = \re^{\ri \l^\dagger} \re^\cV \re^{-\ri \l} ~.
\eea
The $\t$-group is completely gauged away in this representation. 

In the chiral representation, the constraint \eqref{BianchiG} 
is solved as follows: 
\bea
\frak{G}=\frac{\ri}{2}\cDB^\a(\re^{-\cV}\cD_\a\re^{\cV})
~.
\eea
The field strength  $\frak{G}$ is no longer Hermitian.
It obeys  the modified reality condition
$\frak{G}^\dagger =\re^{\cV} \frak{G}  \re^{-\cV}$. 

In the remainder of this section, 
we will work in the chiral representation.

\subsection{$\cN=2$ Chern-Simons and SYM actions}
When dealing with the  non-Abelian $\cN=2$  vector supermultiplet,
it is useful to introduce a covariant variation, $\D \cV$, of the prepotential $\cV$
following \cite{GGRS}. 
It is  defined by 
\bea
\D \cV:=\re^{-\cV}\d \re^{\cV}
~,
\label{varVnA}
\eea
and hence $\d \re^{\cV}=\re^{\cV}\D \cV$
and $\d \re^{-\cV}=-\D \cV\re^{-\cV}$.
Varying the field strength gives
\begin{subequations}
\bea
\d \frak{G} = {[}\frak{G},\D \cV{]}
+\frac{\ri}{2}\cDB^\a\cD_\a\D \cV
-\frac{\ri}{2}\{\cDB^\a\D \cV,\re^{-\cV}\cD_\a \re^{\cV}\}
~,
\label{varGnA}
\eea
which is equivalent to 
\bea
\d \frak{G}
=
\frac{\ri}{2}\bar{\frak{D}}^\a\frak{D}_\a\D \cV
=\frac{\ri}{2}{\frak{D}}^\a\bar{\frak{D}}_\a\D \cV
~.
\eea
\end{subequations}

The $\cN=2$ SYM action in (2,0) AdS superspace is 
a minimal extension of the one in Minkowski space \cite{ZP}, 
\bea
S^{(2,0)}_{\rm SYM} &=& 
-\frac{1}{2g^2}\int \rd^3x\, \rd^2\q  \rd^2\qb\, E \, 
{\rm tr}
{[}\frak{G}^2{]}~,
\eea
with $g$ the coupling constant. 
Its variation is given by 
\bea
\d S^{(2,0)}_{\rm SYM} &=& 
-\frac{\ri}{2g^2}\int \rd^3x\, \rd^2\q  \rd^2\qb\, E \, 
{\rm tr}
\big{[}\D \cV \,\bar{\frak{D}}^\a\frak{D}_\a \frak{G}\big{]}
~.
\eea

We now turn to introducing a supersymmetric Chern-Simons (SCS) action 
in (2,0) AdS superspace.
In the case of Poincar\'e supersymmetry, 
the $\cN=2$  SCS action was constructed by 
Zupnik and Pak \cite{ZP}, and a few years later by Ivanov \cite{Ivanov91}
in a more general form.
Here we follow Ivanov's construction. 
Let us consider a one-parameter family of superfields $\cV(t)$, with $t\in[0,1]$,
 such that
 $\cV(0)=0$ and $\cV(1)=\cV$.
Up to an overall constant, the SCS action is 
\bea
S^{(2,0)}_{\rm SCS}
=  
\int_0^1\rd t
\int\rd^{3}x\,\rd^2\q\rd^2\qb\, E\,{\rm tr}\Big{[}
\frak{G}(t)
\re^{-\cV(t)}\pa_t \re^{\cV(t)}
\Big{]}
~. \label{D.21}
\eea
In the Abelian case this reduces to
\bea
S^{(2,0)}_{\text{SCS-Abelian}}=
\hf \int\rd^{3}x\,\rd^2\q\rd^2\qb\, E\,{\rm tr}
\big{[}
\cV \frak{G}
\big{]}
~.
\eea
Zupnik and Pak \cite{ZP} used the specific parametrization, $\cV(t) = t \cV$.  

It follows from the definition \eqref{varVnA} that 
\bea
\d (\re^{- \cV(t)}\pa_t\re^{\cV(t)})
=
{[}\re^{-\cV(t)}\pa_t\re^{\cV(t)},\D \cV(t){]}
+\pa_t \D \cV(t)
~.
\eea
Making also use of 
\eqref{varGnA}, we compute the variation of the SCS action \eqref{D.21} to be 
\bea
\d S^{(2,0)}_{\rm SCS}
=
\int\rd^{3}x\,\rd^2\q\rd^2\qb\, E\,{\rm tr}\big{[}
\D \cV \,\frak{G}
\big{]}
~.
\eea

\subsection{$\cN=4$ SYM theory with (4,0) AdS supersymmetry}

We are now in a position to present $\cN=2$ superspace formulations
for  all $\cN=4$ SYM actions
which correspond to the different types of $\cN=4$ AdS supersymmetry.
The $R$-charge of $\bm \F$ is universally defined by $\cJ {\bm \F} = - \bm q \bm \F$, 
where ${\bm q}=1+\frac{X}{2S}$. 
In the cases of 
(3,1) and (2,2) AdS supersymmetries, $X$ is equal to zero and $\bm q =1$.

In the case of  (4,0) AdS supersymmetry with  $\bm q \neq 0$, 
the non-manifest supersymmetry transformations are
\bsubeq \label{D.24}
\bea
\d_\ve {\bm \F}
&=&
\ri\big(\ve^\a\bar{\frak{D}}_\a -4S\ve_\rL\big)\frak{G}
=-\frac{1}{2(2-{\bm q})}\bar{\frak{D}}^2(\bar{\ve}_\rR\frak{G})
~,
\label{40tFnA}
\\
\D_\ve \cV
&=&
-\frac{2\ri}{{\bm q}}( \bar{\ve}_\rL {\bm \F} - \ve_\rL \bar {\bm \F}) 
\label{40tVnA}
~, 
\\
\d_\ve \frak{G}
&=&
-\ri\big(\bar{\ve}^\a\frak{D}_\a
-8 S\bar{\ve}_\rL\big){\bm \F}
+{\rm h.c.}
=
\ri \frak{D}_\a(\bar{\ve}^\a{\bm \F})
+{\rm h.c.}
=\frac{\ri}{2}\frak{D}^\a\bar{\frak{D}}_\a\D_\ve\cV
~.
\label{40tGnA}
\eea
\esubeq
These transformation laws are 
non-Abelian extensions of   \eqref{7.1}.

The $\cN=4$ SYM action is 
\bea
S^{(4,0)}_{\rm SYM} &=& \frac{1}{g^2}
\int \rd^3x\, \rd^2\q  \rd^2\qb\, E \, 
{\rm tr}
\Big{[}
{\bar{\bm \F}}{\bm \F}
-\frac{1}{2} \frak{G}^2
+2 S \bm q 
\int_0^1\rd t\,
\frak{G}(t)
\re^{-\cV(t)}\pa_t \re^{\cV(t)}
\Big{]}~.~~ 
\label{D.25}
\eea
It is invariant under the transformations \eqref{D.24}.
The action reduces to \eqref{7.2} in the Abelian limit. 

Our $\cN=4$ SYM action \eqref{D.25} is analogous to the one given in the Euclidean case in \cite{SS}. There is, 
however, a minor technical difference.
The point  is that $\cV (t)$ was chosen in \cite{SS} to be of 
 the form $\cV (t) = t \cV$. In our approach $\cV (t)$ is an arbitrary function  
modulo the boundary conditions $\cV (0) =0$ and $\cV (1) = \cV$ .

In the case of critical (4,0) AdS supersymmetry with
$X+2S= 2S{\bm q}=0$, we have  $\ve_\rL=0$ and 
the non-manifest supersymmetry transformations are
\bsubeq \label{D.277}
\bea
\d_\ve {\bm \F}
&=&
\ri\ve^\a\bar{\frak{D}}_\a \frak{G}
=-\frac{1}{4}\bar{\frak{D}}^2(\bar{\ve}_\rR \frak{G})
~,
\\
\D_\ve \cV
&=&
-2\ri\r\big({\bm\F}-{\bm\Fb}\big)
~,
\\
\d_\ve \frak{G}
&=&
-\ri\bar{\ve}^\a{\frak{D}}_\a{\bm \F}
+{\rm h.c.}
=
\ri {\frak{D}}_\a(\bar{\ve}^\a{\bm \F})
+{\rm h.c.}
=\frac{\ri}{2}\frak{D}^\a\bar{\frak{D}}_\a\D_\ve\cV
~.
\eea
\esubeq
These transformation laws are non-Abelian extensions of \eqref{7.3}.  
The corresponding $\cN=4$ SYM action is given by \eqref{D.25} 
with ${\bm q}=0$. It is an instructive exercise to show that the action is 
invariant under \eqref{D.277}.

\subsection{$\cN=4$ SYM theory with (3,1) AdS supersymmetry}

In the case of (3,1) AdS supersymmetry, the 
non-manifest supersymmetry transformations  
are 
\bsubeq \label{D.27}
\bea
\d_\ve \bm \F
&=&
\ri\big(\ve^\a\bar{\frak{D}}_\a -4S\ve\big) \frak{G}
=
-\frac{1}{2}\bar{\frak{D}}^2\big(
(\bar{\ve}-\bar{\r}) \frak{G}
\big)
~,
\\
\D_\ve \cV
&=&
-2\ri(\bar{\ve}+\bar{\r}) \bm \F +2\ri({\ve}+{\r}) \bar {\bm \F}
~,
\\
\d_\ve \frak{G}
&=&
\ri \frak{D}_\a(\bar{\ve}^\a \bm \F)
+{\rm h.c.}
=\frac{\ri}{2}\frak{D}^\a\bar{\frak{D}}_\a\D_\ve\cV
~.
\eea
\esubeq
These transformation laws are non-Abelian extensions of \eqref{7.4}

The $\cN=4 $ SYM action is
\bea
S^{(3,1)}_{\rm SYM} &=&  \frac{1}{g^2}
\int \rd^3x\, \rd^2\q  \rd^2\qb\, E \, {\rm tr}\Big{[}
\bar {\bm \F} {\bm \F}
-\frac{1}{2} \frak{G}^2
+ S
\int_0^1\rd t\,
\frak{G}(t)
\re^{-\cV(t)}\pa_t \re^{\cV(t)}
\Big{]}
\non\\
&&
- \frac{S}{g^2}\Big{\{} \frac{\ri}{2}  \int \rd^3x\, \rd^2\q   { \cE}\, {\rm tr}[\F^2]
+{\rm c.c.}~\Big{\}}
~.~~~~~~~~~
\eea
It is invariant under \eqref{D.27} and
 reduces to \eqref{7.5} in the Abelian limit.

\subsection{$\cN=4$ SYM theory with (2,2) AdS supersymmetry}

In the case of (2,2) AdS supersymmetry, 
the non-manifest  supersymmetry transformations of $\bm \F$ and $\frak{G}$ are 
\bsubeq \label{D.29}
\bea
\d_\ve \bm \F
&=&
\ri\ve^\a\bar{\frak{D}}_\a \frak{G}
=\frac{1}{2}\bar{\frak{D}}^2\big(
\bar{\r} \frak{G}
\big)
~,
\\
\d_\ve\frak{G}
&=&
-\ri\bar{\ve}^\a\frak{D}_\a \bm \F
+{\rm h.c.}
=\frak{D}^\a\bar{\frak{D}}_\a\big(
\bar{\r}\,{\bm\F}-\r\, \bar {\bm\F }
\big)
~,
\eea
\esubeq
where the parameter $\r$ is defined by \eqref{6.37}.
These transformation laws are non-Abelian extensions of \eqref{7.6}.

The $\cN=4$ SYM  action is
\bea
S^{(2,2)}_{\rm SYM} &=&  \frac{1}{g^2}
\int \rd^3x\, \rd^2\q  \rd^2\qb\, E \, {\rm tr}\Big{[}
\bar {\bm \F} \bm \F
-\frac{1}{2} \frak{G}^2
\Big{]}
~.
\eea
It is invariant under the transformations \eqref{D.29}.

\section{Relating the $\cN=4$ and $\cN=2$ superspace formulations for 
$\cN=4$ SYM theories in AdS$_3$}
\setcounter{equation}{0}

In appendix \ref{AppendixA}, we described the projective-superspace formulation 
for the $\cN=4$ SYM multiplet in a curved superspace of $\cN=4$ supergravity. 
Here we will specify the background superspace to be one of the  $\cN=4 $
AdS superspaces and show how to reduce the formulation of appendix 
\ref{AppendixA} to (2,0) AdS superspace. 
Using the results obtained, we will integrate the variation \eqref{4.8} 
in (2,0) AdS superspace. 

\subsection{Relating the bridge superfields}

In the $\cN=4$ SYM formulation given in appendix  \ref{AppendixA}, 
the fundamental role is played by the bridge 
$\O_+$, eqs. \eqref{A.7} and \eqref{A.8}, and its smile-conjugate $\O_-$ defined by  \eqref{A.9}. 
It is possible to represent 
\bea
\re^{\O_+ (\z_\rR) }=\re^{\O_0}\re^{\hat{\O}_+(\z_\rR)}
~,~~~~~~
\re^{\O_-(\z_\rR)}=\re^{\hat{\O}_-(\z_\rR)}\re^{\O_0^\dagger}
~,
\eea
where 
\bea
\hat{\O}_+(\z_\rR)  = \sum_{n=1}^{\infty} (\z_\rR)^n{\hat\O}_n ~,~~~~~~
\hat{\O}_-(\z_\rR)  = \sum_{n=1}^{\infty} (-1)^n\frac{1}{(\z_\rR)^{n}}{\hat\O}^\dagger_n ~.
\eea
It may be shown that ${\hat \O}_1$, the leading coefficient in the Taylor  
series for $\hat{\O}_+(\z_\rR) $,  is related to $\O_1$, 
the next-to-leading in the Taylor  
series for ${\O}_+(\z_\rR) $, as follows:
\bea
\hat{\O}_1 = \int_0^1 \rd \t \re^{-\t \O_0} \O_1 \re^{\t \O_0}~.
\eea
The gauge covariant spinor derivative ${\frak D}_\a^{(\bar 1) i}$ defined
by eq.~\eqref{A.6}
may be represented as ${\frak D}_\a^{(\bar 1) i} = v^{\bar 1} \frak{D}_\a^{[\bau]i}$, where
$\frak{D}_\a^{[\bau]i}=\re^{\O_0}\re^{\hat{\O}_+}\cD_\a^{[\bau]i}\re^{-\hat{\O}_+}\re^{-\O_0}$ 
is such that
\bea
\frak{D}_\a^{[\bau]i}
=
\frak{D}_\a^{i\bad}
-\z_\rR\frak{D}_\a^{i\bau}
=
\re^{\O_0}\re^{\hat{\O}_+ (\z_\rR)}\cD_\a^{i\bad}\re^{-\hat{\O}_+(\z_\rR)}\re^{-\O_0}
-\z_\rR\re^{\O_0}\re^{\hat{\O}_+(\z_\rR)}\cD_\a^{i\bau}\re^{-\hat{\O}_+(\z_\rR)}\re^{-\O_0}
~.
\eea
We see that the $\z_\rR$-independent part of $\frak{D}_\a^{[\bau]i}$ is
\bea
\frak{D}_\a^{i\bad}
&=&
\re^{\O_0}\cD_\a^{i\bad}\re^{-\O_0}
~.
\eea
Choosing here $i =2$ and projecting to (2,0) AdS superspace gives
\bea
-\frak{D}_\a^{2\bad}|
&=&
-\re^{\O_0|}\cD_\a^{2\bad}|\re^{-\O_0|}
=
\re^{\O}\cDB_\a\re^{-\O}
=\bar{\frak{D}}_\a
~,~~~~~~\O:=\O_0|
~.
\label{E.5}
\eea
Here $\bar{\frak{D}}_\a$ denotes one of the $\cN=2$ gauge covariant derivative, 
eq.~\eqref{D.6}.
Taking the Hermitian conjugate of \eqref{E.5} leads to 
\bea
\frak{D}_\a^{1\bau}|=\frak{D}_\a=\re^{-\O^\dagger}\cD_\a\re^{\O^\dagger}~.
\eea

\subsection{Relating the SYM field strengths} 

Our next task is to reduce the $\cN=4 $ SYM field strength to 
(2,0) AdS superspace.
Making use of \eqref{A.33}, it may be shown that 
\bea
\frak{W}^{ij}=
\re^{\O_0}\,\frak{W}_+^{ij}(\z_\rR=0)\,\re^{-\O_0}
= 
-\frac{1}{4} \re^{\O_0}
\Big{[}\big( 
{\cD}^{ij\bad\bad} 
-4\ri\cS^{ij}{}^{\bad\bad}
\big)
\hat{\O}_1\Big{]}
\re^{-\O_0}
~.~~~~~~~~~~
\label{W222}
\eea
It is convenient to represent the bar-projection of $\frak{W}^{ij}$ 
in terms of the left projective superfield
$\frak{W}^{(2)}(v_\rL) = \frak{W}^{ij}v_iv_j$, where $v^i = v^1(1, \z_\rL)$. 
It follows that 
\bea
\frak{W}^{(2)}(v_\rL)|
&=&
\ri\z_\rL(v^{1})^2\frak{W}^{[2]}(\z_\rL)|~, \qquad
\frak{W}^{[2]}(\z_\rL)|
=
-\frac{\ri}{\z_\rL}{\bm\F}
+\frak{G}
-\ri\z_\rL{\bm\Fb}
~,~~~~~~
\eea
where we have introduced the following $\cN=2$ superfields:
\begin{subequations}
\bea
{\bm\F}&:=&\frak{W}^{22}|~,\qquad \bar{\frak{D}}_\a{\bm\F}=0~, \\
\frak{G}&:=& 2\ri \,\frak{W}^{12}|~,\qquad 
{\frak{D}}^2\frak{G}= 0~, \quad \bar{\frak{D}}^2\frak{G}=0~, \\
{\bm\Fb}&=&\frak{W}^{11}|~, \qquad\frak{D}_\a{\bm\Fb}=0~.
\eea
\end{subequations}
The constraints on $\bm \F$ and $\frak G$ are direct consequences of the 
Bianchi identity obeyed by ${\frak W}^{ij}$.
Since the reduction to (2,0) AdS superspace is characterized by the conditions 
\eqref{condONcS}, from \eqref{W222} we deduce that 
\bea
{\bm\F}=
\frac{1}{4} \re^{\O}
(\cDB^2
\hat{\O}_1|)
\re^{-\O}
=
\frac{1}{4}\bar{\frak{D}}^2{\bm\cX}
~,~~~~~~~~~
{\bm\cX}:=
 \re^{\O}\hat{\O}_1|\re^{-\O}
 ~.
\eea
This is the non-Abelian extension of the first expression in \eqref{6.21}.

Let us now express the covariantly real linear superfield
$\frak{G}:=2\ri \,{\frak W}^{12}|$ in terms of prepotentials.
In this case it is simpler to work in the $\cN=2$ chiral representation 
defined by eqs.~(\ref{D11})--(\ref{D14}).
Using \eqref{W222}, a short calculation gives
\bea
\re^{-\O}\frak{G}\re^{\O}
=
\frac{\ri}{2}  \cDB^\g\cD_\g^{1\bad} \hat{\O}_1|
~.
\label{E11}
\eea
Note that the $\cN=4$ analyticity condition 
$0=\cD^{[1]i}_\a\re^{V(\z_\rR)}=(-\z_\rR\cD^{i\bau}_\a+\cD^{i\bad}_\a)\re^{V(\z_\rR)}$
implies the following constraint on $\O_+$ and $\O_-$:
\bea
\re^{-\O_-}(\cD_\a^{1\bad}\re^{\O_-})
-\z_\rR\re^{-\O_-}(\cD_\a^{1\bau}\re^{\O_-})
=
\re^{\O_+}(\cD_\a^{1\bad}\re^{-\O_+})
-\z_\rR\re^{\O_+}(\cD_\a^{1\bau}\re^{-\O_+})
~.
\label{E.12}
\eea
Picking the linear in $\z_\rR$ term in the Laurent expansion of \eqref{E.12} 
and then bar-projecting to 
(2,0) AdS superspace, we obtain the constraint
\bea
\cD_\a^{1\bad}\hat{\O}_1|
=
\re^{-\cV}\cD_\a\re^{\cV}
~.
\label{E.13}
\eea
Here the right-hand side is expressed in terms of the
$\cN=2$ SYM  prepotential $\cV$ defined by $\re^{\cV}=\re^{\O^\dagger}\re^{\O}$.
Now we can make use of \eqref{E.13}  in \eqref{E11} to obtain
\bea
\re^{-\O}\frak{G}\re^{\O}
=\frac{\ri}{2}  \cDB^\g\big(\re^{-\cV}\cD_\g\re^{\cV}\big)
~,
\eea
which is the $\cN=2$ SYM  field strength in the chiral representation.

\subsection{Integrating the variation of the SYM action}

To conclude this appendix, let us consider
the (2,0) AdS reduction of 
the variation of the $\cN=4$ SYM action, 
eq.~\eqref{4.8}.
The bar-projection of \eqref{4.8} turns out to be
\bea
\d S_{\rm SYM}
 &=& \int \rd^3x\, \rd^2{\q} \rd^2{\qb}\, {  E}\, 
\oint_C \frac{\rd \zeta_\rR}{2\pi \ri \zeta_\rR}\, 
{\rm tr}
\Big{\{}
\Big{[}
\re^{\O_0}\re^{\hat{\O}_+} \D \hat{\O}_+  \re^{-\hat{\O}_+}\re^{-\O_0}
+\re^{-\O_0^\dagger}\re^{-\hat{\O}_-} \D\hat{\O}_- \re^{\hat{\O}_-} \re^{\O_0^\dagger}
\non\\
&&~~~~~~~~~~~~~~~~~~~~~~~~~~~~~~~~~~~~~
+\re^{\O_0}\D\cV\re^{-\O_0}
\Big{]}
{\bm W}^{[{\bar 2}]} 
 \Big{\}}\big|
~,~~~~~~~~~~
\label{E.15}
\eea
where we have defined
\bsubeq
\bea
\D\hat{\O}_+&:=&\re^{-\hat{\O}_+} \d \re^{\hat{\O}_+}
~,\\
\D\hat{\O}_-&:=&(\d\re^{\hat{\O}_-})\re^{-\hat{\O}_-}
~,
\\
\D\cV&=:&\re^{-\cV}\d\re^{\cV}
=\re^{-\O}\d\re^{\O}
+\re^{-\cV}(\d\re^{\O^\dagger})\re^{-\O^\dagger} \re^{\cV}
~.
\eea
\esubeq
We recall that the non-Abelian composite superfield ${\bm W}^{(\bar 2)}$ is defined by 
\eqref{4.9}. The superfield ${\bm W}^{[{\bar 2}]}$ in \eqref{E.15} is 
\bea
{\bm W}^{[\bar 2]}(\z_\rR)&=&
-\frac{\ri}{2\z_\rR}{\bm W}^{\bad\bad}
+2\ri{\bm W}^{\bau\bad}
-\ri\z_\rR{\bm W}^{\bau\bau}
~.
\eea
Computing the bar-projection of the superfields on the right gives 
\bsubeq
\bea
{\bm W}^{\bau\bau}|
&=&
-\frac{\ri}{4}\frak{D}^2 {\bm\F}
+\cS^{22}{}^{\bau \bau }{\bm\Fb}
~,~~~
{\bm W}^{\bad\bad}|=
\frac{\ri}{4}\bar{\frak{D}}^2 {\bm\Fb}
+\cS^{11}{}^{\bad \bad }{\bm\F}
~,~~~~~~
\\
{\bm W}^{\bau\bad}|
&=&
-\frac{1}{4}\big{(}
\frak{D}^{\a}\bar{\frak{D}}_\a
+4\ri \bm q\cS
\big{)} \frak{G}
~.
\eea
\esubeq
Upon evaluation of 
the contour integral in \eqref{E.15} we derive
\bea
\d S_{\rm SYM}
 &=&
  \int \rd^3x\, \rd^2{\q} \rd^2{\qb}\, {  E}\,
{\rm tr}
\Big{\{}\,
\frac{1}{4}\d {\bm\cX}\bar{\frak{D}}^2{\bm\Fb}
+\frac{1}{4}\d{\bm\cX}^\dagger\frak{D}^2 {\bm\F}
-\re^{\O}\D \cV\re^{-\O}\Big(\frac{\ri}{2}\frak{D}^{\a}\bar{\frak{D}}_\a \frak{G}\Big)
\non\\
&&~~~~~~~~~~~~~~~~~~~~~~~~~
+2\cS\bm q\re^{\O}\D \cV\re^{-\O} \frak{G}
-\ri\cS^{11}{}^{\bad \bad }\d {\bm\cX}{\bm\F}
+\ri\cS^{22}{}^{\bau \bau }\d{\bm\cX}^\dagger{\bm\Fb}
 \Big{\}}
~.~~~~~~~~~~~~
\eea
It may be seen that this variation is generated by the action 
\bea
S_{\rm SYM} &=&
\int \rd^3x\, \rd^2\q  \rd^2\qb\, E \, {\rm tr}\Big{[}
\bar {\bm \F} {\bm \F}
-\frac{1}{2} \frak{G}^2
+ 2\cS \bm q 
\int_0^1\rd t\,
\frak{G}(t)
\re^{-\cV(t)}\pa_t \re^{\cV(t)}
\Big{]}
\non\\
&&
+\Big{\{} \frac{\ri}{2} \cS^{11}{}^{\bad \bad } \int \rd^3x\, \rd^2\q   { \cE}\, {\rm tr}[{\bm\F}^2]
+{\rm c.c.}~\Big{\}}
~.~~~~~~~~~
\label{E.21}
\eea
This is indeed the correct action for all the types of $\cN=4$ AdS supersymmetry, 
as discussed
in the previous appendix.
Action \eqref{E.21} is the non-Abelian extension of \eqref{4gen}.


\begin{footnotesize}

\end{footnotesize}

\end{document}